\documentclass[letterpaper,twocolumn,unpublished]{quantumarticle} 
\pdfoutput=1
\usepackage{amsmath,amsthm,amsfonts,amssymb} 

\usepackage{graphicx}
\usepackage{dcolumn}
\usepackage{bm}

\usepackage{tikz}
\usepackage{pgfplots}
\usetikzlibrary{shapes}
\usetikzlibrary{quantikz}
\usetikzlibrary{positioning,arrows,shapes,shadows,calc,fit,backgrounds}
\usepgfplotslibrary{fillbetween}
\usetikzlibrary{positioning}
\pgfplotsset{compat=1.17}

\usepackage{multirow}
\usepackage[ruled,vlined]{algorithm2e}

\usepackage{hyperref}
\hypersetup{  colorlinks=true, linkcolor=blue, citecolor=red, urlcolor=blue  }

\usepackage{color}
\newcommand*{\red}{\textcolor{red}} 
\newcommand*{\bl}{\textcolor{black}} 

\newcommand{\cT}{\mathcal{T}}
\newcommand{\cK}{\mathcal{K}}

\usepackage[caption=false]{subfig}

\begin{document}

\renewcommand{\braket}[1]{\langle{#1}\rangle}

\newcommand{\Op}{\mathbf{Op}}

\newcommand{\ZZ}{\mathbb{Z}}
\newcommand{\NN}{\mathbb{N}}

\newcommand{\cfree}{red!50!black}
\newcommand{\ph}[1]{\phase[red]{#1}}
\newcommand{\phf}[1]{\phase[\cfree,label position=above]{#1}}
\newcommand{\gnum}[1]{\gategroup[1,steps=1, style={color=white, inner ysep=0pt}, label style={yshift=-0.15cm}, background]{{\color{blue} #1}}}
\newcommand{\gnumf}[1]{\gategroup[1,steps=1, style={color=blue, inner sep=-0.2cm}, label style={yshift=0.0cm}]{{\color{blue} #1}}}

\newcommand{\ngate}[2]{\gate{#1}\gategroup[1,steps=1, style={color=white, inner ysep=0pt}, label style={yshift=-0.15cm}, background]{{\color{blue} #2}}}
\newcommand{\nbgate}[2]{\gate{#1}\gategroup[1,steps=1, style={color=white, inner ysep=0pt}, label style={label position=below,anchor=north,yshift=-0.0cm}, background]{{\color{blue} #2}}}
\newcommand{\nctrl}[2]{\ctrl{#1}\gategroup[1+#1,steps=1, style={color=blue, inner sep=-0.15cm}, label style={yshift=-0.05cm}, background]{{\color{blue} #2}}}

\newcommand{\leaftb}[3]{$\begin{array}{c}\color{blue}#1\\\hline\color{red}#2\color{\cfree}#3\end{array}$}

\title{Multi-Tensor Contraction for XEB Verification of Quantum Circuits}%

\author{Gleb Kalachev}
 \email{kalachev@intsys.msu.ru}
\author{Pavel Panteleev}%
 \email{panteleev@intsys.msu.ru}
 \affiliation{Huawei 2012 Lab}%
 \affiliation{Lomonosov Moscow State University.}
\author{Man-Hong Yung}%
 \email{yung.manhong@huawei.com}
\affiliation{Huawei 2012 Lab}%
\affiliation{Institute for Quantum Science and Engineering, and Department of Physics, Southern University of Science and Technology, Shenzhen, 518055, China}%

\begin{abstract}
 The computational advantage of noisy quantum computers has been demonstrated by sampling the bitstrings of quantum random circuits. An~important issue is how the performance of quantum devices could be quantified in the so-called ``supremacy regime''. The~standard approach is through the linear cross entropy benchmark (XEB), where the~theoretical value of the probability is required for each bitstring. However, the~computational cost of XEB grows exponentially. So far, random circuits of the 53-qubit Sycamore chip were verified up to $10$ cycles of gates only; the~XEB fidelities of deeper circuits were approximated with simplified circuits instead. Here we present a~multi-tensor contraction algorithm for speeding up the calculations of XEB for quantum circuits, where the~computational cost can be significantly reduced through some form of memoization. As~a~demonstration, we analyzed the~experimental data of the~53-qubit Sycamore chip and obtained the~exact values of the~corresponding XEB fidelities up to 16 cycles using only moderate computing resources (few GPUs).
 If~the~algorithm was implemented on the~Summit supercomputer, we estimate that for the~supremacy (20 cycles) circuits, it would only cost 7.5 days, which is several orders of magnitude lower than previously estimated in the~literature.
\end{abstract}

\maketitle

\section{Introduction}

Quantum computational supremacy~\cite{Preskill:Supremacy:2012, Aaronson:Supremacy:2016,Yung2018NSR} represents the status where a~universal quantum computing device can accomplish a certain well-defined computational task much faster than any classical computer\footnote{Let us note that there exists another popular approach to quantum supremacy called \emph{boson sampling}~\cite{Aaronson:Boson:2011,Zhong:Boson:2020,Yung2016}, which is not considered in this paper.}. In~2019 Google's quantum team claimed that this goal was achieved~\cite{Supremacy:2019}; their 53-qubit Sycamore  superconducting chip produced one million samples per $200$ seconds with fidelity up to $0.2\%$ from some random quantum circuits with depth $20$, while the same task of random circuit sampling with a~classical supercomputer was predicted by the Google team to require as many as 10,000 years.

Afterwards, many attempts have been made in order to challenge~\cite{Yung2017,IBM:Supremacy:2019} Google's claim by simulating the~same quantum circuits with classical computers~\cite{Hyper-optimized:2021, Alibaba:2020,Napp:2020, Pan:2021}.  
The~most popular approach so far is based on tensor network (TN) contractions; it is an~important tool for classical simulations of large quantum systems~\cite{Markov:TN:2008}, especially when the size of classical memory fails to cover the whole quantum state. Currently, state-of-the-art tensor network algorithms are often applied to estimate the~expectation \mbox{values} of quantum observables~\cite{Alibaba:QAOA:2019}, and evaluate single amplitudes or batches (i.e., a~collection of bitstrings that share some fixed bits) of amplitudes for quantum circuits~\cite{IBM:Batch:2020,Google:TN:2018, Hyper-optimized:2021, Alibaba:2020}. The amplitudes can be obtained by directly contracting all indices in the TN, or by using \emph{slicing}, also called \emph{variables projection}~\cite{Chen:2018, Hyper-optimized:2021, Supremacy:2019, Vincent:2022}. The latter is usually less efficient but reduces the~required memory size and allows to perform the~contraction in parallel. 

In the context of random circuit $C$ simulation with TNs, previous works \cite{Google:TN:2018, IBM:Batch:2020, Schutski:2020, Hyper-optimized:2021, Alibaba:2020, Vincent:2022}  focused mostly on the efficiency in the evaluation of a~single amplitude/probability $p_{C}(s)=\left|\left\langle s|C| 0^{n}\right\rangle\right|^{2}$ or one batch of amplitudes. For example, in~\cite{IBM:Batch:2020} a~batch of size $2^{37}$ is calculated for a~universal random circuit of depth $23$ in a~2D lattice of $8\times 7$ qubits.  Furthermore, the idea of using large batches in quantum simulations as a~trade-off between the~single-amplitude and the~full-state simulators is discussed in~\cite{Schutski:2020}. 

The computational cost of calculating \emph{one}~batch of size~$s$ using TN contractions is usually much smaller than the cost of calculating $s$ independent amplitudes. This significant cost reduction is because when we calculate the amplitudes in a~batch we have a~lot of common subexpressions that can be shared and reused during the contraction algorithm. It was shown very recently~~\cite{Vincent:2022} that this general idea can also be used to reduce the cost of slicing.  

However, in order to perform a full classical simulation of a~quantum circuit~$C$, or to verify the fidelity of the experimental output, one must also consider the problem on how \emph{multiple} uncorrelated (batches of) amplitudes can be evaluated efficiently. Particularly, in Google's experiment~\cite{Supremacy:2019}
the~linear cross-entropy benchmarking (Linear XEB) was proposed as a~tool for estimating the~fidelity of random circuits. Explicitly, the~linear XEB fidelity  $\mathcal{F}_{\mathrm{XEB}}$ for a~sequence of bitstrings $s_{1}, \ldots, s_{k}$, produced by the experiment is defined as
\begin{equation}
\mathcal{F}_{\mathrm{XEB}}\equiv \frac{2^n}{k} \sum_{i=1}^k p_C(s_i) - 1 \ .  
\end{equation}

In other words, for the~\emph{verification task} in random-circuit sampling, one needs to find
the (theoretical) exact amplitudes for the random bitstrings produced in the experiments. At first sight, we may try to minimize the cost of calculations by choosing the batches covering as many as possible the experimental bitstrings. However, the problem is that the sampling size is too small, $k\ll 2^n$ (${\sim}10^{6}$ vs $2^{53}$ in Google's experiment~\cite{Supremacy:2019}), compared with the whole Hilbert space; one would often need to calculate almost all $k$ batches of amplitudes in practice. \bl{Apart from the~verification task, we may also benefit~\cite{Kalachev:2021a} from finding multiple batches of amplitudes if we want to sample from a~quantum circuit $C$ according to its output probability distribution $p_C(s)$ using the~\emph{frugal rejection sampling} method~\cite{Markov:Supremacy:2018, Supremacy:2019}.}

On the other hand, a recent work~\cite{Pan:2021} demonstrated spoofing of the~Linear XEB test in the~aforementioned Google's experiment for the~``supremacy circuit" ($53$ qubits, $20$ cycles), with a~single batch of amplitudes. Here \emph{spoofing} means that, instead of running the~actual simulation with a~classical computer (i.e., output bitstrings according to the distribution of the actual quantum circuit), one produces bitstrings in a way just for passing the statistical test---the~Linear XEB. We should note that despite the~big difference between the~simulation and spoofing tasks for random quantum circuits, the latter is also considered by some researchers to be a~classically-hard problem~\cite{Aaronson:Spoofing:2020}, but in some special cases there exist polynomial-time algorithms~\cite{barak:spoofing:2021}. 

\bl{
In the~current work, we develop a new set of tools for solving problems involving contraction of multiple tensor networks.
%
The~main feature of our approach is to assign a~contraction tree~\cite{Bienstock:1990, OGorman:2019, Hyper-optimized:2021} the \emph{contraction expression}, where pre-calculated sub-expressions are invoked as much as possible. Moreover, a~global cache is utilized to collect these values for speeding up multiple tensor contraction of different (batches of) amplitudes.  As a result, this approach allows us to reduce the~total computational cost by several orders of magnitude, compared to independent multiple runs of the~tensor contraction. Furthermore, this approach is compatible with different TN  contraction algorithms available in the~literature~\cite{Hyper-optimized:2021, Alibaba:2020,Pan:2021}. Here our contraction algorithm is based on local transformations of contraction trees described in Appendix~\ref{sc:cont-tree-opt}. }

 The proposed algorithm was applied to verify the XEB fidelity of the (ABCD) supremacy circuits containing non-simplifiable tiling and sequence of quantum gates~\cite{Supremacy-data:2019}, where no more than $10$ cycles of gates have been verified so far. For this reason, the Google team relied on simplified circuits (elided and patch) to indirectly estimate the~Linear XEB of the supremacy circuits at higher depths. 

Here, with our multi-tensor contraction algorithm, we have successfully verified all ABCD supremacy circuits with $12$, $14$, and $16$ cycles using only moderate computing resources (few GPUs).
Based on our results, we conclude that Google's estimated XEB (based on simplified circuits) contains about $4\%$ deviation. The data produced by our algorithm is available online~\cite{verification-data:2021}.
If our algorithm was implemented on the Summit supercomputer, $16$ cycles would only take 10 mins. Furthermore, we estimate that for verifying the 3 million bitstrings from the 20-cycles supremacy circuits, it would only require 7.5 days, which is several orders of magnitude lower than previously estimated (e.g. $79$ years with the  approach in Ref.~\cite{Alibaba:2020}). 

The~rest of the paper is organized as follows. First, we briefly recall some standard definitions and notations related to tensor networks. Second, we discuss contraction trees and how to estimate their computational cost. \mbox{After} that, we describe our~new efficient method for finding multiple amplitudes and batches of amplitudes, which can be used in combination with any TN~based contraction algorithm. Finally, we demonstrate our experimental results on the~verification task.

\section{Definitions and notations}

To get started, let us summarize the related concepts in TNs necessary for our discussion. Here a~\emph{tensor} of \emph{order} $r$ is a~multi-dimensional array $T[i_1,\dots,i_r] \equiv T[\mathbf{i}]$ with complex entries, where the~indices $(i_1,\dots,i_r) \equiv \mathbf{i}$ are usually called \emph{legs}, and the dimension of each leg is called its \emph{bond dimension}. The~\emph{shape} of the~tensor $T[i_1,\dots,i_r]$ is the~vector $(d_1,\dots,d_r)$, where each $d_j$ is the~bond dimension of the~tensor leg $i_j$;  $j=\overline{1,r}$. For example, a~vector $T[i_1]$ of length $n$ is an~order~$1$ tensor of shape~$(n)$, and an~$m\times n$ matrix $T[i_1,i_2]$ is an~order~$2$ tensor of shape~$(m, n)$.

Later, we would be interested in evaluating the summation of a collection of tensors $T_1[\mathbf{i}_1],\dots, T_m[\mathbf{i}_m]$ sharing some common legs,
\begin{equation}\label{eq:sum}
    {\rm{sum}}=\sum_{j_1,\dots,j_s} T_1[\mathbf{i}_1]\cdots T_m[\mathbf{i}_m],
\end{equation}
where the~sum is over all possible values of the~legs $j_1,\dots,j_s$, which we call the \emph{closed} legs. All the~rest legs of the~tensors $T_1,\dots,T_m$ are called \emph{open}.
As one can see, a~\emph{tensor network} is equivalent to the graphical representation of the summation. 

\begin{figure}
    \centering
    \def\x{0.7cm}
    \subfloat[]{
    \begin{tikzpicture}[
        scale=2,
        tensor/.style={draw,thick,minimum size=15pt,inner sep=0pt,color=black},
        leg/.style={draw,thick,black,-}]
        \node[tensor] (T) at (0,0) {$T$};
        \node[tensor] (S) at (0,-\x) {$S$};
        \node[tensor] (U) at (\x,-0.5*\x) {$U$};
        \node[tensor] (R) at (2*\x,0) {$R$};
        \node[tensor] (Q) at (2*\x,-\x) {$Q$};
        \node[\cfree] (n) at (2.6*\x,0) {};
        \path[leg] (T) -- node[left] {$i$} (S);
        \path[leg] (T) -- node[above] {$j$} (U);
        \path[leg] (S) -- node[below] {$k$} (U);
        \path[leg] (U) -- node[below] {$m$} (Q);
        \path[leg] (R) -- node[right] {$l$} (Q);
        \path[leg] (R) -- node[right] {$l$} (Q);
        \path[leg,red] (R) -- node[below] {$n$} (n);
    \end{tikzpicture}
    }
    \subfloat[]{
    \begin{tikzpicture}[
    point/.style={circle,inner sep=0pt,minimum size=3pt,fill}, 
    level/.style={sibling distance = 5.5em/#1,
    level distance = 1.8em},every node/.style = {align=center}]]
        \node[point] (root) {}
        child { node[point] {} 
          child { node[point] {}
            child { node {$T$}}
            child { node {$U$}}}
          child { node {$S$}}}
        child { node[point] {}
          child { node {$R$}}
          child { node {$Q$}}};
    \end{tikzpicture}
    }
    \caption{(a) Example tensor network; (b) the contraction tree for the~expression
$((T*U)*S)*(R*Q)$.}
    \label{fg:tn-ex}
\end{figure}
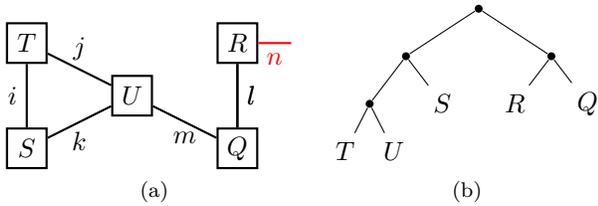

Formally, it can be represented by a~hypergraph $\mathcal{N}$, where each~tensor is denoted by a~vertex, each~leg is denoted by a~hyperedge connecting all the related tensors. We call the sum~(\ref{eq:sum}) the~\emph{result of contraction} for $\mathcal{N}$ denoted by $\Sigma\mathcal{N}$. Furthermore, we also need to specify a~subset $\Op(\mathcal{N})$ of open legs.  For example, in~Fig.~\ref{fg:tn-ex}(a) we have a tensor network $\mathcal{N}$ that corresponds to the~following sum:
\begin{equation}
  \sum_{i,j,k,l,m} T[i,j]S[i,k]U[j,k,m]Q[m,l]R[l,n] \ ,
\end{equation}
where the~open leg~$n$ (shown in red) is not involved in the~summation; therefore $\Op(\mathcal{N}) = \{n\}$. Note that tensor networks can be also viewed as factor graphs, which are widely used in the~context of error-correcting codes and statistical inference~\cite{Factor_graph:2001, Factor_graph:2017}.

\section{Tensor network contraction}

For simplicity, in what follows, we shall consider only tensor networks that are represented by graphs, i.e. each leg connects \emph{at most two} tensors, and it is open whenever it connects to only one tensor. However, all the~algorithms described below also work for arbitrary tensor networks. Suppose we have a pair of tensors $T[i_{1},\ldots,i_{n}]$ and
$S[j_{1},\ldots,j_{m}]$ in a~tensor network $\mathcal{N}$ and a total of $q$ common closed legs in the set. We can define their \emph{contraction} denoted by
$T*_{\mathcal{N}}\text{S~}$ as follows:
\begin{equation}\label{eq:elem-cont}
T*_{\mathcal{N}}S \equiv \sum_{{\rm{closed \ legs}}} T[i_{1},\ldots,i_{n}]\cdot S[j_{1},\ldots,j_{m}].
\end{equation}
In other words, the~contraction of two tensors corresponds
to merging the~corresponding vertices in the~tensor network.
Note that we would omit the index $\mathcal{N}$ if the~tensor network is clear
from the~context and just write $T*S$.

It is not hard to see that if a~tensor network $\mathcal{N}$ consists of tensors
$T_{1},\ldots,T_{n}$, then the result of its contraction $\Sigma\mathcal{N}$ does not depend on the~way we order the~tensors and use parentheses.
For example, for the tensor network from Fig.~\ref{fg:tn-ex}(a) we can use the~following expression:
\begin{equation}\label{eq:contract-ex}
\Sigma\mathcal{N} = ((T*U)*S)*(R*Q) \ .
\end{equation}

The~same result can be obtained by any other expression
that calculates $\Sigma N$, for example: 
\[\Sigma N = ((Q*T)*(S*U))*R.\]
However, from a~practical point of view, a~different contraction expression usually has different computational cost. This cost may be measured in the~number 
of arithmetic floating-point operations such as addition and
multiplication (FLOPs) and the~number of tensor elements we read and
write. Different contraction expressions also have different memory
budgets. We can estimate from below the~required memory size by the~maximal size of
intermediate results (i.e. the~size of intermediate contractions) during
the~evaluation of the~contraction expression. See Appendix~\ref{sc:cont-tree-opt} for more details on the~contraction cost and memory size.

Each contraction expression can be naturally represented by a~binary
tree that is usually called the~\emph{contraction tree}~\cite{Bienstock:1990, OGorman:2019, Hyper-optimized:2021}. In this tree, the~leaves
correspond to the~tensors from the expression and the~internal nodes to the~contractions. 
For example, the tree in Fig.~\ref{fg:tn-ex}(b) corresponds to expression~(\ref{eq:contract-ex}).

\section{Multi-batch simulator}
Before we proceed to our main algorithm, let us first introduce the concept of the contraction of multiple tensor networks. 
Note that both the computational complexity and the memory budget do not depend
on the content of the tensors in the~contraction tree. In fact, they only depend on the~bond dimensions of the~tensors $T_{1}, \ldots, T_{m}$. Thus, it is helpful to consider formal expressions, where instead of some fixed tensors in the contraction expression we have variables $X_{1}, \ldots, X_{m}$ that denote arbitrary tensors of the~same shapes as the tensors $T_{1}, \ldots, T_{m}$. 

We can consider a~contraction tree $\mathcal{T}$ with $m$ leaves also as a~formal contraction expression $\mathcal{T}(X_{1}, \ldots, X_{m})$. Hence we see that the contraction tree $\mathcal{T}$ is just a~pictorial way to represent a~formal contraction expression $\mathcal{T}(X_{1}, \ldots, X_{m})$. Moreover, the subtrees $\mathcal{T}^{\prime}$ of the~contraction tree $\mathcal{T}$ for a~contraction expression $\mathcal{T}(X_{1}, \ldots, X_{m})$ represent its subexpressions $\mathcal{T}^{\prime}(X_{p}, \ldots, X_{q})$. Hence in what follows we are going to identify formal contraction expressions and the~corresponding contraction trees.

By a~\emph{tensor network diagram} we mean a~tensor network $D$, where instead of fixed tensors $T_{1}, \ldots, T_{m}$ of some shapes we have variables $X_{1}, \ldots, X_{m}$ that correspond to arbitrary tensors of the same shapes. If we want to emphasize its variables, we denote a tensor network diagram as $D(X_{1}, \ldots, X_{m}) .$ If we assign tensors $T_{1}, \ldots, T_{m}$ to the~variables $X_{1}, \ldots, X_{m}$ we obtain the~tensor network that we denote by $D(T_{1}, \ldots, T_{m})$. The~result of the~contraction for this tensor network is denoted as $\Sigma D(T_{1}, \ldots, T_{m})$. If $\mathcal{T}(X)$ is a~contraction expression for $D$, then we can use it to perform this contraction, and obtain the~result $\Sigma D(T_{1}, \ldots, T_{m}) = \mathcal{T}(T_{1}, \ldots, T_{m})$. 

In the literature, there are a~number of algorithms~\cite{Hyper-optimized:2021, Alibaba:2020,Pan:2021} for optimizing contraction trees. In~Appendix~\ref{sc:cont-tree-opt} we present our own optimization algorithm used to find contraction trees in this work. In all our experiments, we use a~C++~implementation of this algorithm together with our own efficient library for TN contractions with GPU support.

In order to find $k$ different
amplitudes (resp., batches), a~common approach is just to run a~single-amplitude
(resp., single-batch) contraction algorithm $k$ times. However, this simple method is not efficient in the~case when we need to find a~large number (say ${\sim}10^{6}$) of uncorrelated amplitudes or batches.

Below, we show that there exists a much more efficient way. If we are given a quantum
circuit $C$, then we can convert it into a tensor network $\mathcal{N}_{C}$ in a standard way (see, for example,~\cite{Schutski:2020}). We also suppose that standard TN simplification techniques like \emph{gate fusion} are already applied~\cite{Smelyanskiy:gate-fusion:2016, Hyper-optimized:2021}. This tensor network $\mathcal{N}_{C}$ has $n$ open legs, where $n$ is the number of qubits in our circuit (each open leg corresponds to one output qubit).

Let $D=D(X)$, $X=(X_{1}, \ldots, X_{m})$, be the tensor network diagram for $\mathcal{N}_{C}$ with tensor variables $X_{1}, \ldots, X_{m}$, and $\mathcal{T}(X)$ is a contraction tree for $D(X)$. As it was already mentioned before, in a~multi-amplitude simulation we find $k$ complex amplitudes $\left\langle s_{i}|C| 0^{n}\right\rangle$ for $k$ bitstrings $s_{1}, \ldots, s_{k} \in\{0,1\}^{n}$. We can obtain this as the~result of the~contractions of $k$ tensor networks $D(T^{1}), \ldots, D(T^{k})$, where each collection of tensors $T^{i}=(T_{1}^{i}, \ldots, T_{m}^{i}), i=\overline{1, k}$, corresponds to one bitstring~$s_{i}$ (we assign its bits to the~output legs of $\mathcal{N}_{C}$). If we have some contraction tree $\mathcal{T}(X)=\mathcal{T}(X_{1}, \ldots, X_{m})$ for $D$, then we can use it to perform the~contractions for our $k$ tensor networks $D(T^{1}), \ldots, D(T^{k})$ and obtain: 
\[
\left\langle s_{i}|C| 0^n\right\rangle=\Sigma D(T^i)=\mathcal{T}(T^i); i=\overline{1, k}.
\] 
If one needs to find multiple batches (each of $2^w$ amplitudes) we proceed in a~similar way, but instead of the~full contraction we do not contract $w$ legs that correspond to the non-fixed positions in each batch.   

\begin{figure}
    \centering
    \begin{tikzpicture}[>=stealth']
        \def\x{7.0ex}
        \node[isosceles triangle,draw,minimum size = 12ex,rotate=90] (T1) at (0,0) {};
        \node[anchor=south] at (T1.apex) {\small $\mathcal{T}(T^1)$};
        \draw [->] ([yshift=-1.5ex]T1.125) node[anchor=north,label={[yshift=-4.1ex]{\small $T^1_1$}}] {} -- (T1.125);	
        \draw [->] ([yshift=-1.5ex]T1.235) node[anchor=north,label={[yshift=-4.1ex]{\small $T^1_m$}}] {} -- (T1.235);	
        \node [anchor=north] at (T1.west) {\tiny ...};
        
        \node at (\x,3ex) {\small $\dots$};
        
        \node[isosceles triangle,draw,minimum size = 12ex,rotate=90] (T2) at (2*\x,0) {};
        \node[anchor=south] at (T2.apex) {\small $\mathcal{T}(T^i)$};
        \draw [->] ([yshift=-1.5ex]T2.125) node[anchor=north] {\small $T^i_1$} -- (T2.125);	
        \draw [->] ([yshift=-1.5ex]T2.235) node[anchor=north] {\small $T^i_m$} -- (T2.235);	
        \draw [->] ([yshift=-1.5ex]T2.150) node[anchor=north,color=red] {\small $T^i_p$} -- (T2.150);	
        \draw [->] ([yshift=-1.5ex]T2.210) node[anchor=north,color=red] {\small $T^i_q$} -- (T2.210);	
        \node [anchor=north] at (T2.135) {\tiny ...};
        \node [anchor=north] at (T2.225) {\tiny ...};
        \node [anchor=north] at (T2.west) {\tiny ...};
        \node [isosceles triangle,draw,minimum size = 5ex,rotate=90,anchor=west, fill=lightgray] (T2a) at (T2.west) {};
        \node[anchor=south, color=lightgray] at ([xshift=-1.5ex,yshift=-1.5ex]T2a.apex) {\small $\mathcal{T}'$};
        \draw[dashed] (T2a.apex) -- (T2.apex);
        
        \node at (3*\x,3ex) {\small $\dots$};
        
        \node[isosceles triangle,draw,minimum size = 12ex,rotate=90] (T3) at (4*\x,0) {};
        \node[anchor=south] at (T3.apex) {\small $\mathcal{T}(T^j)$};
        \draw [->] ([yshift=-1.5ex]T3.125) node[anchor=north] {\small $T^j_1$} -- (T3.125);	
        \draw [->] ([yshift=-1.5ex]T3.235) node[anchor=north] {\small $T^j_m$} -- (T3.235);	
        \draw [->] ([yshift=-1.5ex]T3.150) node[anchor=north,color=red] {\small $T^j_p$} -- (T3.150);	
        \draw [->] ([yshift=-1.5ex]T3.210) node[anchor=north,color=red] {\small $T^j_q$} -- (T3.210);	
        \node [anchor=north] at (T3.135) {\tiny ...};
        \node [anchor=north] at (T3.225) {\tiny ...};
        \node [anchor=north] at (T3.west) {\tiny ...};
        \node [isosceles triangle,draw,minimum size = 5ex,rotate=90,anchor=west, fill=lightgray] (T3a) at (T3.west) {};
        \node[anchor=south, color=lightgray] at ([xshift=-1.5ex,yshift=-1.5ex]T3a.apex) {\small $\mathcal{T}'$};
        \draw[dashed] (T3a.apex) -- (T3.apex);

        \node at (5*\x,3ex) {\small $\dots$};

        \node[isosceles triangle,draw,minimum size = 12ex,rotate=90] (T4) at (6*\x,0) {};
        \node[anchor=south] at (T4.apex) {\small $\mathcal{T}(T^k)$};
        \draw [->] ([yshift=-1.5ex]T4.125) node[anchor=north,label={[yshift=-4.1ex]{\small $T^k_1$}}] {} -- (T4.125);	
        \draw [->] ([yshift=-1.5ex]T4.235) node[anchor=north,label={[yshift=-4.1ex]{\small $T^k_m$}}] {} -- (T4.235);	
        \node [anchor=north] at (T4.west) {\tiny ...};

        \node [anchor=north] at (3*\x,-10ex) {\small $T^i_p = T^j_p\!, \dots, T^i_q = T^j_q \ \Longrightarrow\  \mathcal{T}'\!(T^i_p\!,\dots,T^i_q) = \mathcal{T}'\!(T^j_p\!,\dots,T^j_q)$};
    \end{tikzpicture}

    \caption{The~main idea of the~multi-tensor contraction: we can \mbox{evaluate} $\mathcal{T}'(X_p,\dots,X_q)$ only once and reuse the~result next time if the~values of variables $X_p,\dots,X_q$ are the same.}
    \label{fg:main}
\end{figure}
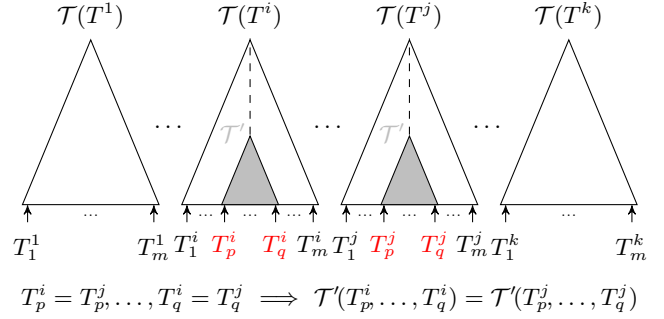

Hence we see that in a~multi-amplitude and multi-batch simulation we evaluate the~contraction expression $\mathcal{T}(X)$ on multiple collections of tensors $T^i=(T_1^i, \ldots, T_m^i)$, $i = \overline{1, k}$. We call this \emph{multi-tensor} contraction procedure since it produces $k$ tensors.
The~key observation is as follows: if one performs these $k$ contractions sequentially
for $i=1,2, \ldots, k$, and we already evaluated some subexpression $\mathcal{T}^{\prime}(X_p,\dots,X_q)$ of
$\mathcal{T}(X)$, then we can reuse the~result next time when the~values of the~variables $X_p,X_{p+1}, \ldots, X_q$ are the same (see Fig.~\ref{fg:main}).

\section{Multi-tensor contraction algorithm}
Below we consider an~algorithm for calculating $k$ contractions $\mathcal{T}(T^1), \ldots, \mathcal{T}(T^k)$ that stores the~intermediate results of all its previous calls in a~global cache~$\mathcal{K}$.  We further assume that $\mathcal{K}$ can be updated while the~algorithm is running. The cache $\mathcal{K}$ can be implemented as a~key lookup data structure. Here the~key is a~tuple $v = (\mathcal{T};T_1,\dots,T_m)$, where $\mathcal{T}=\mathcal{T}(X_1,\dots,X_m)$ is a~contraction expression and $T_1,\dots,T_m$ are the values of its variables $X_1,\dots,X_m$.  The~value $\mathcal{K}(v)$ of the~cache~$\mathcal{K}$, corresponding to the~key~$v$, is equal to the~result $\mathcal{T}(T_1,\dots,T_m)$ of the~expession $\mathcal{T}$ evaluation on $T_1,\dots,T_m$. We also write $\mathcal{K}(v)=\textbf{null}$ if at the current stage we do not have the~entry for the~key~$v$ in the~global cache~$\mathcal{K}$. 

Algorithm~\ref{al:main} shows the~top level procedure that finds $\mathcal{T}(T^1), \ldots, \mathcal{T}(T^k)$ for multiple collections of tensors $T^{i}=(T_1^i, \ldots, T_m^i)$, $i=\overline{1,k}$.
We see that in this procedure we call $k$ times the subprocedure $\mathbf{eval}(\mathcal{T}, T, \mathcal{K})$, which, given the~contraction tree $\mathcal{T}$, a~collection of tensors $T=(T_1,\dots,T_m)$, and the~intermediate results of the~previous calls saved in $\mathcal{K}$, gives us $\mathcal{T}(T)$. Algorithm~\ref{al:eval} shows a~recursive definition of this subprocedure.

\begin{algorithm}[t]
 $\mathcal{K} := \varnothing$ (start with the empty global cache)\;
 \For{$i := 1$ \KwTo $k$}{
    Calculate $\mathcal{T}(T^i):=\mathbf{eval}(\mathcal{T}, T^{i}, \mathcal{K})$\;
  }
 \Return $\mathcal{T}(T^1), \ldots, \mathcal{T}(T^k)$\;
 \caption{Multi-tensor contraction}
 \label{al:main}
\end{algorithm}

\bl{
If we used this algorithm directly, then the~cache size would be very big. However, one can significantly reduce it by reordering the~collections of tensors $T^1,\dots,T^k$ in some~special way, and deleting every~cache entry $\mathcal{K}(v)$ immediately after the~corresponding tensor was used for the~last time. Let us describe how to achieve this. We assume that the~variables $X_1,...,X_m$ from the~top-level contraction expression $\mathcal{T}(X_1,\dots,X_m)$ are enumerated according to their positions in $\mathcal{T}$. We also want to emphasize that each collection of tensors $T^{i}=(T_1^i, \ldots, T_m^i)$ corresponds to an~assignment of values to the~variables $X_1,...,X_m$. Since we have $k$ such collections each variable takes at most $k$ different values, which we can enumerate for each $X_j$, $j=\overline{1,m}$.   This allows us to put $T^1,\dots,T^k$ in the~lexicographic order. To reduce the size of the~cache $\mathcal{K}$ it can be split into the~left and right parts $\mathcal{K}_L$ and $\mathcal{K}_R$ for storing the~results of the~left and right subexpressions in Algorithm~\ref{al:eval}, respectively. This splitting allows us to store in the~left cache $\mathcal{K}_L$ \emph{at most one} entry for each subexpression; and before we store $\mathcal{K}_L(\mathcal{T};T_1,...,T_m)$, we can remove all keys $(\mathcal{T};...)$ from $\mathcal{K}_L$. The lexicographic ordering guarantees that the~removed keys will not be used anymore.
}

\bl{
To obtain a~close-to-optimal contraction cost during the~multi-tensor contraction we need to find a~good contraction expression $\mathcal{T}$. The main characteristics that should be considered here are as follows:
\begin{enumerate}
\item
   \emph{Memory budget} $\mathbf{M} = \mathbf{M}(\mathcal{T)}$, i.e., the~amount of memory required for the~simulation, including the cache size and memory for intermediate
  contraction results;
\item
  \emph{Computational complexity} $\mathbf{C} = \mathbf{C}(\mathcal{T)}$, i.e., 
  the~number of floating-point operations (FLOPs), calculated as the~sum of the~complexities 
  of all contractions in the~contraction expression  $\mathcal{T}$;
\item
   Parameter $\mathbf{RW} = \mathbf{RW}(\mathcal{T)}$, which is equal to the~\emph{number 
  of read-write operations} from the~memory for all contractions
  in the~contraction expression $\mathcal{T}$.
\end{enumerate}
The~parameters $\mathbf{C}$ and $\mathbf{RW}$ should take into account how many times each subexpression is calculated in the~worst case when we perform a~multi-tensor contraction. For example, in the case of multi-amplitude simulation the~complexity may depend on the~number of calculated amplitudes. If we calculate $2^m$ amplitudes, and all the~tensors in a~subexpression $\mathcal{T}$ contain $r$ legs corresponding to the~circuit output, then this subexpression will be evaluated at most $2^{\min(r,m)}$ times. Some further details on the~contraction expression optimization can be found in Appendices~\ref{sc:cont-tree-opt}, \ref{sc:exapmple}, and \ref{sc:details}.
}

\begin{algorithm}[t]
 \lIf{$\mathcal{T}=X_{j}$}{\Return $T_{j}$}
  Let $X_{i_{1}}, \ldots, X_{i_{s}}$ be the variables of $\mathcal{T}$\; 
  \If{$\mathcal{K}(\mathcal{T}; T_{i_{1}}, \ldots, T_{i_{s}})=\mathbf{null}$}{
  Let $\mathcal{T}=\mathcal{T}_L*\mathcal{T}_R$\;
  \tcp{Recursively call itself on subtrees}
  ${U_{L}:=\mathbf{eval}(\mathcal{T}_{L}, T, \mathcal{K})}$\; 
  ${U_{R}:=\mathbf{eval}(\mathcal{T}_{R}, T, \mathcal{K})}$\;
  \tcp{perform the contraction operation}
  $U:=U_{L} * U_{R}$\;
  \tcp{store the result $U$ to the cache $\mathcal{K}$}
  $\mathcal{K}(\mathcal{T}; T_{i_{1}}, \ldots, T_{i_{s}}):=U$\; 
}
\Return $\mathcal{K}(\mathcal{T}; T_{i_{1}}, \ldots, T_{i_{s}})$\;
\caption{Procedure $\mathbf{eval}(\mathcal{T}, T, \mathcal{K})$}
\label{al:eval}
\end{algorithm}

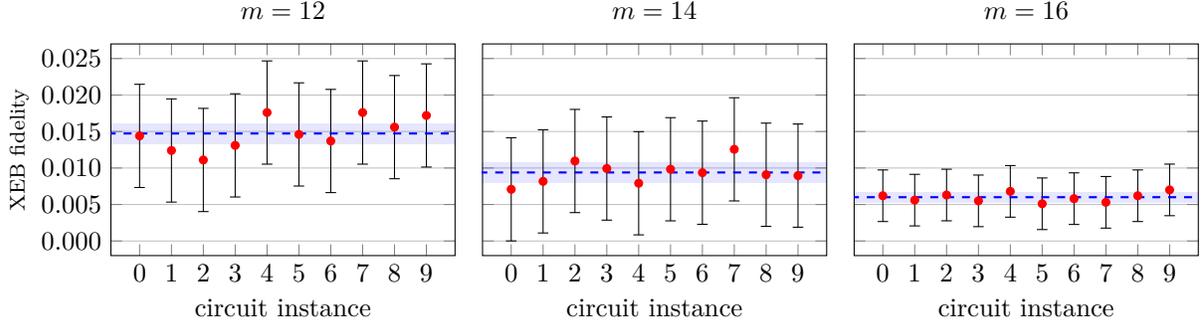
\begin{figure*}[t]
    \centering
    \begin{tikzpicture}
    \begin{axis}[
      yticklabel style={scaled ticks=false,
        /pgf/number format/fixed,
        /pgf/number format/fixed zerofill={true},
        /pgf/number format/precision=3},
      xtick={0,1,...,9},ytick={0,0.005,...,0.025},ymin=-0.002, ymax=0.027, 
      ymajorgrids=true,
      major grid style={line width=.2pt,draw=gray!50},
      xlabel=circuit instance,
      ylabel={XEB fidelity},
      width=0.35\linewidth,
      height=0.25\linewidth,
      only marks,
      every axis y label/.style=
       {at={([xshift=-5pt]ticklabel cs:0.5)},rotate=90,anchor=center,font=\footnotesize},
      title={$m=12$},
      legend pos=north west,
      legend style={draw=none}]
        \draw [name path=upper,draw=none,fill=blue,opacity=0.1] (axis cs:-1,0.01473+0.00141) -- (axis cs:10,0.01473+0.00141) -- (axis cs:10,0.01473-0.00141) -- (axis cs:-1,0.01473-0.00141);
        \addplot+[
          black, mark options={red, scale=0.75},
          error bars/.cd, 
            y fixed,
            y dir=both, 
            y explicit
        ] table [x=x, y=y,y error=error, col sep=comma] {
            x,      y,        error
            0,      0.0144,   0.0070711
            1,      0.0124,   0.0070711
            2,      0.0111,   0.0070711
            3,      0.0131,   0.0070711
            4,      0.0176,   0.0070711
            5,      0.0146,   0.0070711
            6,      0.0137,   0.0070711
            7,      0.0176,   0.0070711
            8,      0.0156,   0.0070711
            9,      0.0172,   0.0070711
        };
        \draw[dashed,blue,thick] (axis cs:-1,0.01473) -- (axis cs:10,0.01473);
    \end{axis}
    \end{tikzpicture}
    \begin{tikzpicture}
    \begin{axis}[
      yticklabel style={scaled ticks=false},
      xtick={0,1,...,9},ytick={0,0.005,...,0.025},ymin=-0.002, ymax=0.027, 
      ymajorgrids=true,
      major grid style={line width=.2pt,draw=gray!50},
      width=0.35\linewidth,
      height=0.25\linewidth,
      yticklabels=\empty,
      xlabel=circuit instance,
      only marks,
      title={$m=14$},
      legend pos=north west,
      legend style={draw=none}]
        \draw [name path=upper,draw=none,fill=blue,opacity=0.1] (axis cs:-1,0.00939+0.00141) -- (axis cs:10,0.00939+0.00141) -- (axis cs:10,0.00939-0.00141) -- (axis cs:-1,0.00939-0.00141);
        \addplot+[
          black, mark options={red, scale=0.75},
          error bars/.cd, 
            y fixed,
            y dir=both, 
            y explicit
        ] table [x=x, y=y,y error=error, col sep=comma] {
            x,      y,        error
            0,   0.00708,   0.0070711
            1,   0.00817,   0.0070711
            2,   0.01096,   0.0070711
            3,   0.00993,   0.0070711
            4,   0.00791,   0.0070711
            5,   0.00983,   0.0070711
            6,   0.00936,   0.0070711
            7,   0.01256,   0.0070711
            8,   0.00908,   0.0070711
            9,   0.00896,   0.0070711
        };
        \draw[dashed,blue,thick] (axis cs:-1,0.00939) -- (axis cs:10,0.00939);
    \end{axis}
    \end{tikzpicture}
    \begin{tikzpicture}
    \begin{axis}[
      yticklabel style={scaled ticks=false},
      xtick={0,1,...,9},ytick={0,0.005,...,0.025},ymin=-0.002, ymax=0.027, 
      ymajorgrids=true,
      major grid style={line width=.2pt,draw=gray!50},
      width=0.35\linewidth,
      height=0.25\linewidth,
      yticklabels=\empty,
      xlabel=circuit instance,
      only marks,
      title={$m=16$},
      legend pos=north west,
      legend style={draw=none}]
        \draw [name path=upper,draw=none,fill=blue,opacity=0.1] (axis cs:-1,0.0060+0.0007071) -- (axis cs:10,0.0060+0.0007071) -- (axis cs:10,0.0060-0.0007071) -- (axis cs:-1,0.0060-0.0007071);
        \addplot+[
          black, mark options={red, scale=0.75},
          error bars/.cd, 
            y fixed,
            y dir=both, 
            y explicit
        ] table [x=x, y=y,y error=error, col sep=comma] {
            x,      y,        error
            0,   0.0062,   0.003535
            1,   0.0056,   0.003535
            2,   0.0063,   0.003535
            3,   0.0055,   0.003535
            4,   0.0068,   0.003535
            5,   0.0051,   0.003535
            6,   0.0058,   0.003535
            7,   0.0053,   0.003535
            8,   0.0062,   0.003535
            9,   0.0070,   0.003535
        };
        \draw[dashed,blue,thick] (axis cs:-1,0.0060) -- (axis cs:10,0.0060);
    \end{axis}
    \end{tikzpicture}
    \caption{The~Linear XEB for all Google's ABCD supremacy circuits for $m=12,14,16$. We show the $\pm 5\sigma$ statistical error bars for each instance and the~band corresponding to $\pm\sigma$ around the~mean fidelity, where $\sigma=1/\sqrt{k}$; $k$ is the~number of samples.}
    \label{fg:XEB}
\end{figure*}

\begingroup 
\begin{table*}
    \centering
    \subfloat[Linear XEB(\%)]{
    \resizebox{9cm}{!}{%
    \begin{tabular}[b]{|c|c|c|c|c|c|c|c|c|c|c|c|c|}
         \hline
         $m$&1&2&3&4&5&6&7&8&9&10&mean &\parbox{1.3cm}{\vphantom{$2^2$}Google's\\ estimate\vphantom{g}\parbox{1.3cm}} \\
         \hline
         \vphantom{\large A}12& 1.44 & 1.24&	1.11&	1.31&	1.76&	1.46&	1.38&	1.76&	1.57&	1.72&	1.47&1.4\\
         \vphantom{$2^2$}14& 0.71 & 0.82 & 1.10 & 0.99&	0.79& 0.98&	0.94&	1.26&	0.91&	0.90&	0.94&0.9\\
         \vphantom{$2^2$}16& 0.62 & 0.56 & 0.63 & 0.55& 0.68& 0.51& 0.58& 0.53& 0.62& 0.70& 0.60&0.6\\
         \hline
    \end{tabular}
    }
    }
    \subfloat[Verification complexity]{
    \resizebox{8cm}{!}{%
    \begin{tabular}[b]{|c|c|c|c|c|c|c|c|c|}
    \hline
        \multirow{2}{*}{$m$}	&\multirow{2}{*}{$k$}	&\multicolumn{2}{c|}{Contraction cost}&	\multicolumn{2}{c|}{Efficiency}&	\multicolumn{3}{c|}{Time (days or years)}\\
        \cline{3-9}
        &&\vphantom{$2^2$}1 amp (S)& $k$ amps (M)& S& M& S & M &gain\\
	\hline
        \vphantom{\large A}12&	0.5M&	$1.8\cdot 10^{13}$&	$2.8\cdot 10^{17}$&	61\%&	43\%&	94 d  &	4.3 d& 22x\\
        14&	0.5M&	$1.0\cdot 10^{14}$&	$1.9\cdot 10^{18}$&	60\%&	60\%&	538 d  &	21 d&25x\\
        16&	2M  &	$8.9\cdot 10^{16}$&	$1.4\cdot 10^{19}$&	63\%&	48\%&	5000 y&	0.5 y&10000x\\
    \hline
    \end{tabular}
    }
    }
    \caption{(a) The~linear XEB(\%) of Google's ABCD supremacy circuits for different number of cycles $m$ = 12, 14, and 16; Google's estimation of XEB is from~\cite[Table~XI, Supplementary Information]{Supremacy:2019}. (b)~The verification complexity of single amplitude (S) and multi-amplitude (M) simulation. The~last column is the~gain of the~multi-amplitude simulator over multiple runs of the~single amplitude simulator. The~time is shown for one Tesla V100 16GB PCI-E. The~number of FLOPs for each case is equal to $8C$, where $C$ is the~contraction cost. The~\mbox{efficiency} here means the~ratio of the~real performance of our implementation to the~peak theoretical performance of a~given GPU.}
    \label{tb:XEB}
\end{table*}
\endgroup

\section{Verification of Google's experiment}
Using the described above multi-amplitude algorithm we verify Google's results~\cite{Supremacy:2019, Supremacy-data:2019} for up to $16$ cycles using the~samples (0.5M--2M samples per circuit) produced in Google’s experiment. We used $4$ identical servers, each with the~following configuration: $2$ GPUs Tesla V100 with 16GB memory, 2 $\times$ Intel(R) Xeon(R) Gold 6151 CPU 3.00GHz. 

A~link to the~archive with the~calculated amplitudes can be found here~\cite{verification-data:2021}.
Based on this data we estimated the~fidelity using the Linear XEB (see Table~\ref{tb:XEB}(a)). In Fig.~\ref{fg:XEB} you can also see these fidelities for $m=12,14,16$ together with the~corresponding mean value and the~standard deviation. As we can see, these results confirm the~fidelities indirectly estimated in Google's paper~\cite{Supremacy:2019, Supremacy-data:2019}.

\begin{figure}[ht]
    \begin{tikzpicture}
    \begin{semilogyaxis}[
      minor grid style={dotted},
      ymin=0.005, ymax=0.1, 
      ymajorgrids=true,
      yminorgrids=true,
      major grid style={line width=.2pt,draw=gray!50},
      width=\linewidth,
      xlabel=number of qubits $n$,
      ylabel=XEB fidelity,
      title={$m=14$ (EFGH)},
      legend pos=north east,
      legend style={cells={anchor=west}}
      ]
        \addplot+[
          black, mark options={black, scale=0.3}, only marks,
        ] table [x=x, y=y, col sep=comma] {
            x,      y
            39,    0.03350
            39,    0.03618
            39,    0.03321
            39,    0.03554
            39,    0.03480
            39,    0.03427
            39,    0.03378
            39,    0.03283
            39,    0.03562
            39,    0.03347
            40,    0.03008
            40,    0.02957
            40,    0.03045
            40,    0.03069
            40,    0.02865
            40,    0.02706
            40,    0.02880
            40,    0.02981
            40,    0.02894
            40,    0.03148
            41,    0.02380
            41,    0.02635
            41,    0.02596
            41,    0.02421
            41,    0.02278
            41,    0.02200
            41,    0.02681
            41,    0.02036
            41,    0.02628
            41,    0.02227
            42,    0.02580
            42,    0.02734
            42,    0.02208
            42,    0.02073
            42,    0.02228
            42,    0.02175
            42,    0.02219
            42,    0.02221
            42,    0.02172
            42,    0.02341
            43,    0.01774
            43,    0.02041
            43,    0.01703
            43,    0.01860
            43,    0.01853
            43,    0.01769
            43,    0.01869
            43,    0.01950
            43,    0.01916
            43,    0.01751
            44,    0.01841
            44,    0.01970
            44,    0.01887
            44,    0.01779
            44,    0.01566
            44,    0.01923
            44,    0.01552
            44,    0.01768
            44,    0.02088
            44,    0.02056
            45,    0.01553
            45,    0.01796
            45,    0.01758
            45,    0.01965
            45,    0.01690
            45,    0.01625
            45,    0.01687
            45,    0.01448
            45,    0.01508
            45,    0.01513
            46,    0.01574
            46,    0.01310
            46,    0.01509
            46,    0.01766
            46,    0.01491
            46,    0.01689
            46,    0.01630
            46,    0.01410
            46,    0.01527
            46,    0.01545
            47,    0.01596
            47,    0.01146
            47,    0.01180
            47,    0.01326
            47,    0.01152
            47,    0.01222
            47,    0.01301
            47,    0.01133
            47,    0.01399
            47,    0.01301
            48,    0.00909
            48,    0.01278
            48,    0.01319
            48,    0.01185
            48,    0.01118
            48,    0.01369
            48,    0.01207
            48,    0.01241
            48,    0.01101
            48,    0.01323
            49,    0.00911
            49,    0.00823
            49,    0.00754
            49,    0.00892
            49,    0.00717
            49,    0.01241
            49,    0.00918
            49,    0.00764
            49,    0.01079
            49,    0.01299
            50,    0.00650
            50,    0.01030
            50,    0.00796
            50,    0.00868
            50,    0.00930
            50,    0.00987
            50,    0.00838
            50,    0.00935
            50,    0.00942
            50,    0.01053
            51,    0.00865
            51,    0.00786
            51,    0.00789
            51,    0.00981
            51,    0.00765
            51,    0.00943
            51,    0.00865
            51,    0.00923
            51,    0.00919
            51,    0.00892        
            };
            \addlegendentry{XEB for 10 instances}
        \addplot+[
          black, mark=*, mark options={red, scale=0.8}, only marks,
        ] table [x=x, y=y, col sep=comma]{
        x,   y
        39,    0.03432
        40,    0.02955
        41,    0.02408
        42,    0.02295
        43,    0.01849
        44,    0.01843
        45,    0.01654
        46,    0.01545
        47,    0.01276
        48,    0.01205
        49,    0.00940
        50,    0.00903
        51,    0.00873
        };
        \addlegendentry{Mean XEB}
        \addplot+[
          black!50!green, mark=+, mark options={black!50!green, scale=1}, 
        ] table [x=x, y=y, col sep=comma]{
        x,   y
        39,0.03073
        40,0.02751
        41,0.02514
        42,0.0229
        43,0.02043
        44,0.01901
        45,0.016
        46,0.01485
        47,0.01281
        48,0.01134
        49,0.01088
        50,0.009337
        51,0.008978
        53,0.007666
        };
        \addlegendentry{Prediction}
        \draw[dashed,blue,thick] (axis cs:-1,0.00939) -- (axis cs:10,0.00939);
    \end{semilogyaxis}
    \end{tikzpicture}
    \caption{The verification results for Google's EFGH circuits with $14$ cycles and the number of qubits $n < 53$. For each $n$ we found the Linear XEB for all 10 EFGH circuits (black dots) and calculated the corresponding mean value (red dots). The green curve shows the predicted XEB values  from~\cite[FIG.~4]{Supremacy:2019}.}
    \label{fg:EFGH}
\end{figure}
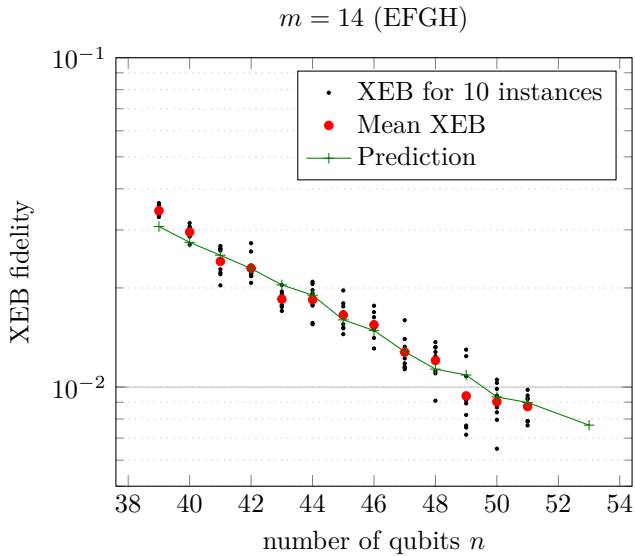

We also verified all the EFGH circuits with $14$ cycles and the number of qubits $n < 53$. The results are shown in Fig.~\ref{fg:EFGH}. We can see that the obtained XEB values (shown in red) are in good correspondence with the theoretical prediction (shown in green) from~\cite[FIG.~4]{Supremacy:2019} (see also~\cite[Eq.~(77), Supplementary Information]{Supremacy:2019}).

The~contraction cost of the~verification task (the~number of arithmetic operations with complex numbers) and its running time on \emph{one} GPU Tesla V100 for single-amplitude (S) and multi-amplitude (M) simulators are shown in Table~\ref{tb:XEB}(b).  For the~single-amplitude case, we assume that the simulator should be run $k$ times to obtain $k$ amplitudes. As we can see,  the~gain of the~multi-amplitude simulator over the~multiple runs of the~single-amplitude one is up to ${~}10^4$ in the~hardest case $m=16$. 

In Table~\ref{tb:ver_compare} we estimated the~hypothetical running time of different algorithms for Summit supercomputer. For qsimh we used the~formula $0.2\cdot 1/f\cdot T_\mathrm{sim}$; where $f$ is the~fidelity, $T_\mathrm{sim}$ is the~running time of the~qsimh simulation on 1M cores with fidelity $f$~\cite[Table~XI, Supplementary Information]{Supremacy:2019}. Here the~factor $0.2$ is because we assume that Summit is approximately equivalent to 5M cores. Let us note that this formula gives a~slightly smaller qsimh running time estimate than the~estimation from~\cite[Fig.~S50, Supplementary Information]{Supremacy:2019}.

In the~future we plan to verify some other cases as well. In~fact, Table~\ref{tb:ver_compare} shows that even in the~case of $m=20$ cycles the~verification of 3M samples can be done in several days on Summit supercomputer.  We should note the~running time of our algorithm depends on the~maximal memory size required during the contraction. In all our estimations we assume that the~GPU memory size is limited by 16GB. This is in a~high contrast with the~well known idea~\cite{IBM:Supremacy:2019} to store all $2^{53}$ amplitudes on hard drives.   We also estimated that for modern GPUs with larger memory sizes, such as Tesla A100 80GB, it is possible to reduce the~running time several times. Moreover, the third generation of tensor cores with better floating point precision, introduced recently in NVIDIA Ampere architecture, can improve the~performance of our algorithm even further.

\begingroup
\begin{table}[t]
    \centering
    \resizebox{8.3cm}{!}{%
    \begin{tabular}{|c|r|r|r|r|r|}
    \hline
        $m$& \#bitstrings & qsimh & Alibaba & Our(S) & Our(M) \\
	\hline
        12&	0.5M & 28 hours     & 11 min    & 5 min      &  14 sec  \\
        14&	0.5M & 300 days     & 73 min    & 28 min     & 1.1 min   \\
        16&	  2M & 133 years    & 348 days  & 66 days    &  10 min   \\
        18& 2.5M & 8,750 years   & 2.2 years & 0.83 years & 1.4 hours \\
        20&   3M & 1,000,000 years & 79 years & 21 years   & 7.5 days  \\
    \hline
    \end{tabular}
    }
    \caption{The estimated time on Summit supercomputer for different simulation algorithms possible to use for the~verification of Google's experiment: Google's hybrid Shr\"odinger-Feynman (SFA) simulator qsimh (multi-amplitude, running time is scaled to 5M CPU cores)~\cite{Supremacy:2019}, Alibaba's simulator~\cite[Table~1]{Alibaba:2020} (single-amplitude), and our TN contraction algorithm for single (S) and multiple (M) amplitudes. We assume that \mbox{Summit} has theoretical 400~PFlop/s single-precision $\sim$ 5M CPU cores with AVX-512. For all single-amplitude simulations the~running time is multiplied by the~number of samples.}
    \label{tb:ver_compare}
\end{table}
\endgroup

\section{Acknowledgments}
We would like to thank Dingshun Lv and Yongqing Liu for their valuable technical help during this~experiment and illuminating discussions. We are also grateful to Pan Zhang and PengFei Zhou for pointing out several misprints and valuable suggestions. We want to express our gratitude to Gil Kalai for a~suggestion to verify the~EFGH circuits of depth~$14$ in Google's experiment.


\bibliography{simulation}

\begin{thebibliography}{10}

\bibitem{Preskill:Supremacy:2012}
John Preskill.
\newblock Quantum computing and the entanglement frontier.
\newblock November 2012.
\newblock \href {http://arxiv.org/abs/1203.5813} {\path{arXiv:1203.5813}}.

\bibitem{Aaronson:Supremacy:2016}
Scott Aaronson and Lijie Chen.
\newblock Complexity-theoretic foundations of quantum supremacy experiments.
\newblock In {\em Proceedings of the 32nd {{Computational Complexity
  Conference}}}, {{CCC}} '17, pages 1--67, {Dagstuhl, DEU}, July 2017. {Schloss
  Dagstuhl--Leibniz-Zentrum fuer Informatik}.

\bibitem{Yung2018NSR}
Man-Hong Yung.
\newblock {Quantum supremacy: some fundamental concepts}.
\newblock {\em National Science Review}, 6(1):22--23, jan 2019.
\newblock \href {https://doi.org/10.1093/nsr/nwy072}
  {\path{doi:10.1093/nsr/nwy072}}.

\bibitem{Aaronson:Boson:2011}
Scott Aaronson and Alex Arkhipov.
\newblock The computational complexity of linear optics.
\newblock In {\em Proceedings of the Forty-Third Annual {{ACM}} Symposium on
  {{Theory}} of Computing}, {{STOC}} '11, pages 333--342, {New York, NY, USA},
  June 2011. {Association for Computing Machinery}.
\newblock \href {https://doi.org/10.1145/1993636.1993682}
  {\path{doi:10.1145/1993636.1993682}}.

\bibitem{Zhong:Boson:2020}
Han-Sen Zhong, Hui Wang, Yu-Hao Deng, Ming-Cheng Chen, Li-Chao Peng, Yi-Han
  Luo, Jian Qin, Dian Wu, Xing Ding, Yi~Hu, Peng Hu, Xiao-Yan Yang, Wei-Jun
  Zhang, Hao Li, Yuxuan Li, and {others}.
\newblock Quantum computational advantage using photons.
\newblock December 2020.
\newblock \href {http://arxiv.org/abs/2012.01625} {\path{arXiv:2012.01625}},
  \href {https://doi.org/10.1126/science.abe8770}
  {\path{doi:10.1126/science.abe8770}}.

\bibitem{Yung2016}
Man-Hong Yung, Xun Gao, and Joonsuk Huh.
\newblock {Universal bound on sampling bosons in linear optics and its
  computational implications}.
\newblock {\em National Science Review}, 6(4):719--729, jul 2019.
\newblock \href {https://doi.org/10.1093/nsr/nwz048}
  {\path{doi:10.1093/nsr/nwz048}}.

\bibitem{Supremacy:2019}
Frank Arute, Kunal Arya, Ryan Babbush, Dave Bacon, Joseph~C. Bardin, Rami
  Barends, Rupak Biswas, Sergio Boixo, Fernando G. S.~L. Brandao, David~A.
  Buell, Brian Burkett, Yu~Chen, Zijun Chen, Ben Chiaro, Roberto Collins, and
  {others}.
\newblock Quantum supremacy using a programmable superconducting processor.
\newblock {\em Nature}, 574(7779):505--510, October 2019.
\newblock \href {https://doi.org/10.1038/s41586-019-1666-5}
  {\path{doi:10.1038/s41586-019-1666-5}}.

\bibitem{Yung2017}
Man-Hong Yung and Xun Gao.
\newblock {Can Chaotic Quantum Circuits Maintain Quantum Supremacy under
  Noise?}
\newblock jun 2017.
\newblock URL: \url{http://arxiv.org/abs/1706.08913}, \href
  {http://arxiv.org/abs/1706.08913} {\path{arXiv:1706.08913}}.

\bibitem{IBM:Supremacy:2019}
Edwin Pednault, John~A. Gunnels, Giacomo Nannicini, Lior Horesh, and Robert
  Wisnieff.
\newblock Leveraging {{Secondary Storage}} to {{Simulate Deep}} 54-qubit
  {{Sycamore Circuits}}.
\newblock October 2019.
\newblock \href {http://arxiv.org/abs/1910.09534} {\path{arXiv:1910.09534}}.

\bibitem{Hyper-optimized:2021}
Johnnie Gray and Stefanos Kourtis.
\newblock Hyper-optimized tensor network contraction.
\newblock {\em Quantum}, 5:410, March 2021.
\newblock \href {https://doi.org/10.22331/q-2021-03-15-410}
  {\path{doi:10.22331/q-2021-03-15-410}}.

\bibitem{Alibaba:2020}
Cupjin Huang, Fang Zhang, Michael Newman, Junjie Cai, Xun Gao, Zhengxiong Tian,
  Junyin Wu, Haihong Xu, Huanjun Yu, Bo~Yuan, Mario Szegedy, Yaoyun Shi, and
  Jianxin Chen.
\newblock Classical {{Simulation}} of {{Quantum Supremacy Circuits}}.
\newblock May 2020.
\newblock \href {http://arxiv.org/abs/2005.06787} {\path{arXiv:2005.06787}}.

\bibitem{Napp:2020}
John Napp, Rolando~L. La~Placa, Alexander~M. Dalzell, Fernando G. S.~L.
  Brandao, and Aram~W. Harrow.
\newblock Efficient classical simulation of random shallow {{2D}} quantum
  circuits.
\newblock March 2020.
\newblock \href {http://arxiv.org/abs/2001.00021} {\path{arXiv:2001.00021}}.

\bibitem{Pan:2021}
Feng Pan and Pan Zhang.
\newblock Simulating the {{Sycamore}} quantum supremacy circuits.
\newblock March 2021.
\newblock \href {http://arxiv.org/abs/2103.03074} {\path{arXiv:2103.03074}}.

\bibitem{Markov:TN:2008}
Igor~L. Markov and Yaoyun Shi.
\newblock Simulating {{Quantum Computation}} by {{Contracting Tensor
  Networks}}.
\newblock {\em SIAM Journal on Computing}, 38(3):963--981, January 2008.
\newblock \href {https://doi.org/10.1137/050644756}
  {\path{doi:10.1137/050644756}}.

\bibitem{Alibaba:QAOA:2019}
Fang Zhang, Cupjin Huang, Michael Newman, Junjie Cai, Huanjun Yu, Zhengxiong
  Tian, Bo~Yuan, Haihong Xu, Junyin Wu, Xun Gao, Jianxin Chen, Mario Szegedy,
  and Yaoyun Shi.
\newblock Alibaba {{Cloud Quantum Development Platform}}: {{Large}}-{{Scale
  Classical Simulation}} of {{Quantum Circuits}}.
\newblock September 2019.
\newblock \href {http://arxiv.org/abs/1907.11217} {\path{arXiv:1907.11217}}.

\bibitem{IBM:Batch:2020}
Edwin Pednault, John~A. Gunnels, Giacomo Nannicini, Lior Horesh, Thomas
  Magerlein, Edgar Solomonik, Erik~W. Draeger, Eric~T. Holland, and Robert
  Wisnieff.
\newblock Pareto-{{Efficient Quantum Circuit Simulation Using Tensor
  Contraction Deferral}}.
\newblock August 2020.
\newblock \href {http://arxiv.org/abs/1710.05867} {\path{arXiv:1710.05867}}.

\bibitem{Google:TN:2018}
Sergio Boixo, Sergei~V. Isakov, Vadim~N. Smelyanskiy, Ryan Babbush, Nan Ding,
  Zhang Jiang, Michael~J. Bremner, John~M. Martinis, and Hartmut Neven.
\newblock Characterizing quantum supremacy in near-term devices.
\newblock {\em Nature Physics}, 14(6):595--600, June 2018.
\newblock \href {http://arxiv.org/abs/1608.00263} {\path{arXiv:1608.00263}},
  \href {https://doi.org/10.1038/s41567-018-0124-x}
  {\path{doi:10.1038/s41567-018-0124-x}}.

\bibitem{Chen:2018}
Jianxin Chen, Fang Zhang, Cupjin Huang, Michael Newman, and Yaoyun Shi.
\newblock Classical {{Simulation}} of {{Intermediate}}-{{Size Quantum
  Circuits}}.
\newblock May 2018.
\newblock \href {http://arxiv.org/abs/1805.01450} {\path{arXiv:1805.01450}}.

\bibitem{Vincent:2022}
Trevor Vincent, Lee~J. O'Riordan, Mikhail Andrenkov, Jack Brown, Nathan
  Killoran, Haoyu Qi, and Ish Dhand.
\newblock Jet: Fast quantum circuit simulations with parallel task-based
  tensor-network contraction.
\newblock {\em Quantum}, 6:709, May 2022.
\newblock URL: \url{https://quantum-journal.org/papers/q-2022-05-09-709/},
  \href {https://doi.org/10.22331/q-2022-05-09-709}
  {\path{doi:10.22331/q-2022-05-09-709}}.

\bibitem{Schutski:2020}
Roman Schutski, Danil Lykov, and Ivan Oseledets.
\newblock Adaptive algorithm for quantum circuit simulation.
\newblock {\em Physical Review A}, 101(4):042335, April 2020.
\newblock \href {http://arxiv.org/abs/1911.12242} {\path{arXiv:1911.12242}},
  \href {https://doi.org/10.1103/PhysRevA.101.042335}
  {\path{doi:10.1103/PhysRevA.101.042335}}.

\bibitem{Kalachev:2021a}
Gleb Kalachev, Pavel Panteleev, PengFei Zhou, and Man-Hong Yung.
\newblock Classical sampling of random quantum circuits with bounded fidelity,
  December 2021.
\newblock URL: \url{http://arxiv.org/abs/2112.15083}, \href
  {http://arxiv.org/abs/2112.15083} {\path{arXiv:2112.15083}}.

\bibitem{Markov:Supremacy:2018}
Igor~L. Markov, Aneeqa Fatima, Sergei~V. Isakov, and Sergio Boixo.
\newblock Quantum {{Supremacy Is Both Closer}} and {{Farther}} than {{It
  Appears}}.
\newblock September 2018.
\newblock \href {http://arxiv.org/abs/1807.10749} {\path{arXiv:1807.10749}}.

\bibitem{Aaronson:Spoofing:2020}
Scott Aaronson and Sam Gunn.
\newblock On the {{Classical Hardness}} of {{Spoofing Linear Cross}}-{{Entropy
  Benchmarking}}.
\newblock {\em Theory of Computing}, 16(11):1--8, November 2020.
\newblock \href {https://doi.org/10.4086/toc.2020.v016a011}
  {\path{doi:10.4086/toc.2020.v016a011}}.

\bibitem{barak:spoofing:2021}
Boaz Barak, Chi-Ning Chou, and Xun Gao.
\newblock {Spoofing Linear Cross-Entropy Benchmarking in Shallow Quantum
  Circuits}.
\newblock In James~R. Lee, editor, {\em 12th Innovations in Theoretical
  Computer Science Conference (ITCS 2021)}, volume 185 of {\em Leibniz
  International Proceedings in Informatics (LIPIcs)}, pages 30:1--30:20,
  Dagstuhl, Germany, 2021. Schloss Dagstuhl--Leibniz-Zentrum f{\"u}r
  Informatik.
\newblock URL: \url{https://drops.dagstuhl.de/opus/volltexte/2021/13569}, \href
  {https://doi.org/10.4230/LIPIcs.ITCS.2021.30}
  {\path{doi:10.4230/LIPIcs.ITCS.2021.30}}.

\bibitem{Bienstock:1990}
Dan Bienstock.
\newblock On embedding graphs in trees.
\newblock {\em Journal of Combinatorial Theory, Series B}, 49(1):103--136, June
  1990.
\newblock \href {https://doi.org/10.1016/0095-8956(90)90066-9}
  {\path{doi:10.1016/0095-8956(90)90066-9}}.

\bibitem{OGorman:2019}
Bryan O'Gorman.
\newblock Parameterization of {{Tensor Network Contraction}}.
\newblock In Wim van Dam and Laura Mancinska, editors, {\em 14th {{Conference}}
  on the {{Theory}} of {{Quantum Computation}}, {{Communication}} and
  {{Cryptography}} ({{TQC}} 2019)}, volume 135 of {\em Leibniz {{International
  Proceedings}} in {{Informatics}} ({{LIPIcs}})}, pages 10:1--10:19, {Dagstuhl,
  Germany}, 2019. {Schloss Dagstuhl\textendash Leibniz-Zentrum fuer
  Informatik}.
\newblock \href {https://doi.org/10.4230/LIPIcs.TQC.2019.10}
  {\path{doi:10.4230/LIPIcs.TQC.2019.10}}.

\bibitem{Supremacy-data:2019}
{The datasets generated in Google's quantum supremacy experiment}.
\newblock \url{https://datadryad.org/stash/dataset/doi:10.5061/dryad.k6t1rj8}.
\newblock [Accessed: 18-February-2020].

\bibitem{verification-data:2021}
{The amplitudes for the samples generated in Google's quantum supremacy
  experiment (multi-volume zip archive)}.
\newblock
  \url{https://gitee.com/Huawei-HiQ/supremacy/tree/master/verification}.
\newblock [Accessed: 25-January-2021].

\bibitem{Factor_graph:2001}
F.R. Kschischang, B.J. Frey, and H.-A. Loeliger.
\newblock Factor graphs and the sum-product algorithm.
\newblock {\em IEEE Transactions on Information Theory}, 47(2):498--519,
  February 2001.
\newblock \href {https://doi.org/10.1109/18.910572}
  {\path{doi:10.1109/18.910572}}.

\bibitem{Factor_graph:2017}
Hans-Andrea Loeliger and Pascal~O. Vontobel.
\newblock Factor {{Graphs}} for {{Quantum Probabilities}}.
\newblock {\em IEEE Transactions on Information Theory}, 63(9):5642--5665,
  September 2017.
\newblock \href {https://doi.org/10.1109/TIT.2017.2716422}
  {\path{doi:10.1109/TIT.2017.2716422}}.

\bibitem{Smelyanskiy:gate-fusion:2016}
Mikhail Smelyanskiy, Nicolas P.~D. Sawaya, and Al{\'a}n {Aspuru-Guzik}.
\newblock {{qHiPSTER}}: {{The Quantum High Performance Software Testing
  Environment}}.
\newblock May 2016.
\newblock \href {http://arxiv.org/abs/1601.07195} {\path{arXiv:1601.07195}}.

\bibitem{Julia}
The contraction order optimizer for {OMEinsum}
  ({TensorBFS/OMEinsumContractionOrders}).
\newblock URL:
  \url{https://juliahub.com/ui/Packages/OMEinsumContractionOrders/xKHvN/0.6.1}.

\bibitem{Williams:arithm-intens:2008}
Samuel~Webb Williams.
\newblock {\em Auto-Tuning Performance on Multicore Computers}.
\newblock PhD thesis, EECS Department, University of California, Berkeley,
  December 2008.

\bibitem{Russell:2009}
Stuart Russell and Peter Norvig.
\newblock {\em Artificial {{Intelligence}}: {{A Modern Approach}}}.
\newblock {Prentice Hall Press}, {USA}, 3rd edition, 2009.

\end{thebibliography}

\appendix
\section{Contraction trees optimization}\label{sc:cont-tree-opt}

In this section we describe a~new tensor contraction algorithm that finds contraction trees using local transformations\footnote{Note that an~open source implementation of the contraction algorithm that closely follows the approach proposed in the current paper can be found in~\cite[the TreeSA method]{Julia}.}.

When a~tensor network is obtained from a~quantum circuit, operating on qubits, all bond dimensions of the~legs are equal to $2$. In this case, the~computational cost~$\mathbf{C}$ of elementary contraction operation~(\ref{eq:elem-cont}) is easier to estimate: it involves $2^{q + r}$ multiplications and almost the~same number of
additions, where $r$ is the~number of
open legs in the result $T*S$. This is because we need to sum up $2^{q}$ terms and do it for all possible $2^{r}$ values of $r$ open legs. 

On the other hand, the~number of memory operations $\mathbf{RW}$ is also an~important
parameter, since read/write operations of the tensors may become the~bottleneck of the~contraction in practice. As an~estimation,
let us consider the~costs of reading the~tensors $T,S$,
and writing the~result $T*S$. The total~number of operations is simply 
 $\text{size}(T) + \text{size}(S) + \text{size}(T*S)$,
where $\text{size}(X)$ is the~\emph{size} of the~tensor $X$, i.e. the~product 
of all bond dimensions of its legs.

It is important to note that for the best overall performance of the
contraction algorithm it is wise to take into account that the memory
speed and the computation speed on particular hardware are not the
same. Hence we need a parameter during our optimization algorithm that
encodes the ratio of these two speeds. This optimization parameter is
called the~\emph{arithmetic intensity}~\cite[Sec.~4.2.2]{Williams:arithm-intens:2008}. 
We define it as the ratio of
computational complexity (the number of elementary floating-point
operations) to the number of memory read/write operations during the
tensor contraction. For example, in GPU Tesla V100 this value is
approximately equal to $16$. Hence the~arithmetic intensity is a~device-dependent parameter, which is different for different hardware.

For optimization, we use the following objective function that tries to combine all the
above characteristics:
\[
f(\mathcal{T}) := \beta \max \left(\log _{2}\left(\frac{\mathbf{M}}{\mathbf{M}_{\max }}\right), 0\right)+\log _{2}(\mathbf{C}+\alpha \cdot \mathbf{RW})
\]
where $\mathbf{M}_{\max }$ is the~upper limit on the~memory size in Bytes (our memory budget); $\alpha$ is the~arithmetic intensity; $\beta$ is the~\emph{penalty factor} for running out of memory, i.e., $\beta$ controls the~weight of memory size in the~objective function. If the~memory budget is more important, we should increase the~value of $\beta$.

To find a~close-to-optimal contraction tree, we need an optimization algorithm that tries to minimize the~objective function $f(\mathcal{T})$. In this work we use \emph{simulated annealing} but any other local search algorithms such as \emph{hill climbing} can be used as well (see~\cite[Chap.~4]{Russell:2009} for a~review of local search methods). By a~\emph{local search method} here we mean a~combinatorial optimization method that given an~objective function $f\colon X \rightarrow \mathbb{R}$ on the~search space $X$ of all possible states try to apply a~small fixed number of local transformations $L=\left\{l_{1}, \ldots, l_{n}\right\}$ (each transformation $l_{i}$ is a
function $l_{i}\colon X \rightarrow X$ ) starting usually from some random or predefined state $x_{0} \in X$. Hence we obtain a sequence of states $x_{0}, x_{1}, \ldots, x_{N}$, where each next state $x_{i+1}$ is obtained from the~previous state $x_{i}$ using one of the~local transformations from the set $L$, i.e. $x_{i+1}=l(x_{i})$ for some $l \in L$. The choice of the local transformation $l \in L$ on each individual step is usually governed by the~\emph{gain}
\[
\Delta f(x_i, l):=f(x_i) - f(l(x_i))
\]
that we obtain in terms of the objective function $f$. The~local search usually stops when
it reaches a~state $x_{N}$ that cannot be improved locally (i.e., $\Delta f(x_{i}, l) < 0$ for all $l \in L$) or the~maximal number of steps is reached. There are many other details on how a~local search can be done. For example, in the~simulated annealing method, we choose local transformations randomly with the~probability that depends on the~gain $\Delta f(x_{i}, l)$. At the same
time, in the~hill-climbing method, one can choose a~local transformation deterministically in
a~greedy fashion (i.e., choose the~local transformation which gives the best possible
gain). 

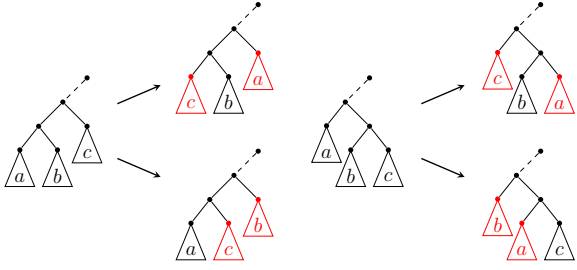
\begin{figure}
    \centering
    \tikzstyle{point}=[circle,inner sep=0pt,minimum size=3pt,fill]
    \tikzstyle{subtree}=[isosceles triangle,draw,minimum size = 4ex,shape border rotate=90,inner sep =-2pt]
    \newcommand{\LeftTree}[6]{%
    \begin{tikzpicture}[node distance=0.7cm]
        \node[point] (N1) {};
        \node[point, below left of = N1] (N2) {};
        \node[point, below left of = N2] (N3) {};
        \node[point, below left = 0.4cm and 0.3cm of N3,color=#4] (N4) {};
        \node[subtree, anchor=apex,color=#4] at (N4) (S1) {\large $#1$};
        \node[point, below right = 0.4cm and 0.3cm of N3,color=#5] (N5) {};
        \node[subtree, anchor=apex,color=#5] at (N5) (S2) {\large $#2$};
        \node[point, below right of = N2,color=#6] (N6) {};
        \node[subtree, anchor=apex,color=#6] at (N6) (S3) {\large $#3$};
    
        \draw[dashed] (N1) -- (N2);
        \draw (N2) -- (N3);
        \draw (N2) -- (N6);
        \draw (N3) -- (N4);
        \draw (N3) -- (N5);
    \end{tikzpicture}}
    \newcommand{\RightTree}[6]{%
    \begin{tikzpicture}[node distance=0.7cm]
        \node[point] (N1) {};
        \node[point, below left of = N1] (N2) {};
        \node[point, below right of = N2] (N3) {};
        \node[point, below left = 0.4cm and 0.3cm of N3,color=#5] (N4) {};
        \node[subtree, anchor=apex,color=#5] at (N4) (S1) {\large $#2$};
        \node[point, below right = 0.4cm and 0.3cm of N3,color=#6] (N5) {};
        \node[subtree, anchor=apex,color=#6] at (N5) (S2) {\large $#3$};
        \node[point, below left = 0.4cm and 0.3cm of  N2,color=#4] (N6) {};
        \node[subtree, anchor=apex,color=#4] at (N6) (S3) {\large $#1$};
        
        \draw[dashed] (N1) -- (N2);
        \draw (N2) -- (N3);
        \draw (N2) -- (N6);
        \draw (N3) -- (N4);
        \draw (N3) -- (N5);
    \end{tikzpicture}}    
    \scalebox{0.65}{
    \begin{tikzpicture}[>=stealth]
        \node[rectangle,minimum width = 2.6cm] (T1) at (0,0) {\LeftTree{a}{b}{c}{black}{black}{black}};
        \node[rectangle,minimum width = 2.6cm] (T2) at (3.5cm,1.5cm) {\LeftTree{c}{b}{a}{red}{black}{red}};
        \node[rectangle,minimum width = 2.6cm] (T3) at (3.5cm,-1.5cm) {\LeftTree{a}{c}{b}{black}{red}{red}};
        
        \draw[->,thick] (T1) -- (T2);
        \draw[->,thick] (T1) -- (T3);
    \end{tikzpicture}
    \begin{tikzpicture}[>=stealth]
        \node[rectangle,minimum width = 2.6cm] (T1) at (0,0) {\RightTree{a}{b}{c}{black}{black}{black}};
        \node[rectangle,minimum width = 2.6cm] (T2) at (3.5cm,1.5cm) {\RightTree{c}{b}{a}{red}{black}{red}};
        \node[rectangle,minimum width = 2.6cm] (T3) at (3.5cm,-1.5cm) {\RightTree{b}{a}{c}{red}{red}{black}};
        
        \draw[->,thick] (T1) -- (T2);
        \draw[->,thick] (T1) -- (T3);
    \end{tikzpicture}
    }
    \caption{Local transformations of contraction trees (triangles correspond to the~subtrees).}
    \label{fg:contr-tree-changes}
\end{figure}

It can be easily checked that the contraction operation $T*S$ (as a
binary operation on tensors) satisfies the~following \emph{associativity}
and \emph{commutativity} conditions:
\begin{itemize}
\item
  $T*(S*R) = (T*S)*R$ \quad (associativity);
\item
  $T*S = S*T$ \quad (commutativity).
\end{itemize}
Using these two conditions, we can deduce the~following identities (see also Fig.~\ref{fg:contr-tree-changes}), which we use as the~local transformations in our~local search algorithm:
 \begin{align*}
   & (a*b)*c \rightarrow (\red{c}*b)*\red{a}, & a*(b*c) \rightarrow \red{c}*(b*\red{a}),\\
   & (a*b)*c \rightarrow (a*\red{c})*\red{b}, & a*(b*c) \rightarrow \red{b}*(\red{a}*c).
 \end{align*}

\begin{figure}
    \centering
    \scalebox{0.53}{
    \begin{tikzpicture}
        \newcommand*{\Bigs}[1]{\vcenter{\hbox{\scalebox{1.5}{\ensuremath#1}}}}
        \tikzstyle{description}=[draw, inner sep=1pt]
        \node[anchor=north west] at (0,0) {$\big((T * U) * S\big) *(R * Q) \rightarrow \big((\red{S} * U) * \red{T}\big) *(R * Q) \rightarrow \red{R} *\Bigs(\!\red{\big((S * U) * T\big)} * Q\Bigs) \rightarrow \red{Q} *\Bigs(\!\big((S * U) * T\big) * \red{R}\Bigs)$};
        \node[description] at (25.6ex,0.9ex) {\scalebox{1.0}{$(a*b)*c \to (\red{c}*b)*\red{a}$}};
        \node[description] at (52.5ex,0.9ex) {\scalebox{1.0}{$a*(b*c) \to \red{b}*(\red{a}*c)$}};
        \node[description] at (80.0ex,0.9ex) {\scalebox{1.0}{$a*(b*c) \to \red{c}*(b*\red{a})$}};
    \end{tikzpicture}
    }
    \caption{An example of local search: on each step we apply one of the~four local transformations.}
    \label{fg:opt-ex}
\end{figure}

\begin{figure}[t]
    \centering
    \def\x{0.6cm}
    \scalebox{0.85}{
    \begin{tikzpicture}[>=stealth,
        scale=1.85,
        tensor/.style={draw,thick,minimum size=15pt,inner sep=0pt,color=black},
        leg/.style={draw,thick,black,-}]
        \node[tensor] (T) at (0,0) {$T$};
        \node[tensor] (S) at (0,-\x) {$S$};
        \node[tensor] (U) at (\x,-0.5*\x) {$U$};
        \node[tensor] (R) at (2*\x,0) {$R$};
        \node[tensor] (Q) at (2*\x,-\x) {$Q$};
        \node (n) at (2.6*\x,0) {};
        \node (L) at (\x,0.5*\x) {sliced variable \red{$v$}};
        \path[leg] (T) -- (S);
        \path[leg,red] (T) -- (U);
        \path[leg] (S) -- (U);
        \path[leg] (U) -- (Q);
        \path[leg] (R) -- (Q);
        \path[leg] (R) -- (Q);
        \path[leg] (R) -- (n);
        \path[name path=LS] (L) -- (S);
        \path[name path=TU] (T) -- (U);
        \path [name intersections={of=LS and TU,by={I}}];
        \draw[->] (L) -- (I);
        
        \node (n1) at (2.85*\x,-0.5*\x) {};
        \node (n2) at (4.5*\x,-0.5*\x) {};
        \draw[->,thick] (n1) edge[above] node {slicing for \red{$v$}} (n2);
        \node (sum) at (5*\x,-0.5*\x) {$\displaystyle\sum_{\red{v\in\{0,1\}}}$};
        
        \node[tensor] (T) at ([xshift=5.75*\x]T) {$T$};
        \node[tensor] (S) at ([xshift=5.75*\x]S) {$S$};
        \node[tensor] (U) at ([xshift=5.75*\x]U) {$U$};
        \node[tensor] (R) at ([xshift=5.75*\x]R) {$R$};
        \node[tensor] (Q) at ([xshift=5.75*\x]Q) {$Q$};
        \node (n) at ([xshift=5.75*\x]n) {};
        \node[inner sep=1pt] (Tn) at ([xshift=0.5*\x]T) {\red{$v$}};
        \node[inner sep=1pt] (Un) at ([xshift=-0.5*\x]U) {\red{$v$}};
        
        \path[leg] (T) -- (S);
        \path[leg] (S) -- (U);
        \path[leg] (U) -- (Q);
        \path[leg] (R) -- (Q);
        \path[leg] (R) -- (Q);
        \path[leg] (R) -- (n);        
        \path[leg,red] (Tn) -- (T);        
        \path[leg,red] (Un) -- (U);        
    \end{tikzpicture}
    }
    \caption{Slicing of the~leg $v$.}
    \label{fg:slicing}
\end{figure}
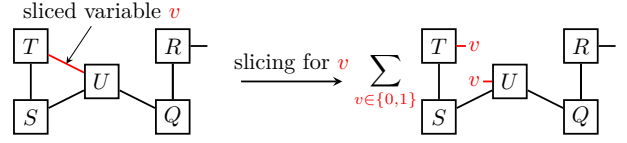

The~set of states in the~local search is the~set
of all possible contraction trees for a~tensor network $\mathcal{N}$, which we also interpret as
contraction expressions we use to find the result of the contraction $\Sigma \mathcal{N}$. We suppose
that on each step of our local search method one of these four local transformations
can be applied to any subexpression (i.e., to a subtree $\mathcal{T}^{\prime}$ of the contraction tree $\mathcal{T}$). In
Fig.~\ref{fg:opt-ex} you can find an~example of some possible steps of our local search
method.

\begin{figure*}[ht]
    \centering
    \subfloat[Quantum circuit] {
    \scalebox{0.62}{
    \begin{quantikz}[baseline=(current bounding box.south)]
        \gate[style={white}]{\ket{0}}\gnumf{$X_0$} & \ph{0} & \ngate{H}{$X_3$} & \ph{3}& \nctrl{1}{$X_5$} &\ph{4}& \qw      & \qw      &\ngate{H}{$X_7$} &\phf{8} \rstick[wires=3]{\color{\cfree}Open \\\color{\cfree}legs}\\
        \gate[style={white}]{\ket{0}}\gnumf{$X_1$} & \ph{1} & \qw      &  \qw     & \targ{}  &\ph{5}& \nctrl{1}{$X_6$} & \ph{7}& \nbgate{H}{$X_8$}& \phf{9} \\
        \gate[style={white}]{\ket{0}}\gnumf{$X_2$} & \ph{2} &\ngate{T}{$X_4$}  &  \qw & \qw     & \ph{6}& \targ{} & \qw & \qw&\phf{10} 
    \end{quantikz}\hspace{2em}
    }}
    \subfloat[Tensor network] {
    \scalebox{0.62}{
    \begin{tikzpicture}[baseline=(current bounding box.south),
        scale=1.2,
        tensor/.style={draw=black,thick,minimum size=15pt,inner sep=0pt,color=blue},
        output/.style={circle,fill=red,minimum size=4pt,inner sep=0pt},
        leg/.style={draw,thick,red,-}]
        \node[tensor] (0) at (0,0) {$X_0$};
        \node[tensor] (3) at (1,0) {$X_3$};
        \node[tensor] (1) at (0,-1) {$X_1$};
        \node[tensor] (2) at (-1,-2) {$X_2$};
        \node[tensor] (4) at (0,-2) {$X_4$};
        \node[tensor] (5) at (1,-1) {$X_5$};
        \node[tensor] (6) at (1,-2) {$X_6$};
        \node[tensor] (7) at (2,-1) {$X_7$};
        \node[tensor] (8) at (2,-2) {$X_8$};
        \node[\cfree] (l8) at (2.7,-1) {8};
        \node[\cfree] (l9) at (2.7,-2) {9};
        \node[\cfree] (l10) at (1,-2.7) {10};
        \path[leg] (0) -- node[above] {0} (3) -- node[right] {3} (5) -- node[above] {4} (7) -- (l8);
        \path[leg] (1) -- node[above] {1} (5) -- node[right] {5} (6) -- node[above] {7} (8) -- (l9);
        \path[leg] (2) -- node[above] {2} (4) -- node[above] {6} (6) -- (l10);
    \end{tikzpicture}
    }}
    \subfloat[Contraction tree] {
    \scalebox{0.62}{
    \begin{tikzpicture}[baseline=(current bounding box.south),
    grow=down,
    level 1/.style=={sibling distance=6cm},
    level/.style={sibling distance = 8cm/(2^(#1-1)), level distance=1.2cm},
    punkt/.style={rectangle, shade, rounded corners, thick, top color=white, draw=black},
    leaf/.style={punkt, inner sep=0pt},
    vleaf/.style={leaf, rectangle split, rectangle split horizontal=true, rectangle split parts=2}
    ]
    \node[punkt, rectangle split, rectangle split horizontal=true, rectangle split parts=2, inner sep=0pt] {\color{\cfree}\ 8,9,10\hphantom{ } \nodepart{second}\color{\cfree}$\tiny\begin{array}{c|c|c}8&9&10\\\hline0&0&0\\1&0&0\\1&1&1\end{array}$}
        child[sibling distance=7cm] {
            node[punkt, rectangle split, rectangle split horizontal=true, rectangle split parts=2, inner sep=0pt] {\color{red}\ 4,\color{red!40!black} 9,10\hphantom{ } \nodepart{second}\color{\cfree}$\tiny\begin{array}{c|c}9&10\\\hline0&0\\1&1\end{array}$}
            child {
                node[punkt] {\color{red}4,5}
                child {
                    node[punkt] {\color{red}3}
                    child{node[leaf] {\leaftb{X_0}{0}{}}}
                    child{node[leaf] {\leaftb{X_3}{0,3}{}}}
                }
                child[sibling distance=2cm] {
                    node[punkt] {\color{red}3,4,5}
                    child{node[leaf] {\leaftb{X_1}{1}{}}}
                    child[sibling distance=1.6cm]{node[leaf] {\leaftb{X_5}{1,3,4,5}{}}}
                }
            }
            child[sibling distance=5.2cm] {
                node[punkt, rectangle split, rectangle split horizontal=true, rectangle split parts=2, inner sep=0pt] {\color{red}\ 5,\color{red!40!black}9,10\hphantom{ }\nodepart{second}\color{\cfree}$\tiny\begin{array}{c|c}9&10\\\hline0&0\\1&1\end{array}$}
                child {
                    node[punkt] {\color{red}6}
                    child{node[leaf] {\leaftb{X_2}{2}{}}}
                    child{node[leaf] {\leaftb{X_4}{2,6}{}}}
                }
                child[sibling distance=4.3cm] {
                    node[punkt, rectangle split, rectangle split horizontal=true, rectangle split parts=2, inner sep=0pt] {\color{red}\ 5,6,\color{red!40!black}9,10\hphantom{ } \nodepart{second}\color{\cfree}$\tiny\begin{array}{c|c}9&10\\\hline0&0\\1&1\end{array}$}
                    child[sibling distance=1.8cm]{
                        node[punkt, rectangle split, rectangle split horizontal=true, rectangle split parts=2, inner sep=0pt] {\leaftb{X_6}{5,6,7,}{10}\nodepart{second}$\color{\cfree}\tiny\begin{array}{c}10\\\hline 0\\1\end{array}$}
                    }
                    child[sibling distance=2.3cm]{
                        node[punkt, rectangle split, rectangle split horizontal=true, rectangle split parts=2, inner sep=0pt] {\leaftb{X_8}{7,}{9}\nodepart{second}$\color{\cfree}\tiny\begin{array}{c}9\\\hline 0\\1\end{array}$}
                    }
                }
            }
        }
        child[sibling distance=6cm] {
            node[punkt, rectangle split, rectangle split horizontal=true, rectangle split parts=2, inner sep=0pt] {\leaftb{X_7}{4,}{8}\nodepart{second}$\color{\cfree}\tiny\begin{array}{c}8\\\hline 0\\1\end{array}$}
         };
    \end{tikzpicture}
    }}
    \caption{Example of multi-amplitude contraction.}
    \label{fg:circuit}
\end{figure*}
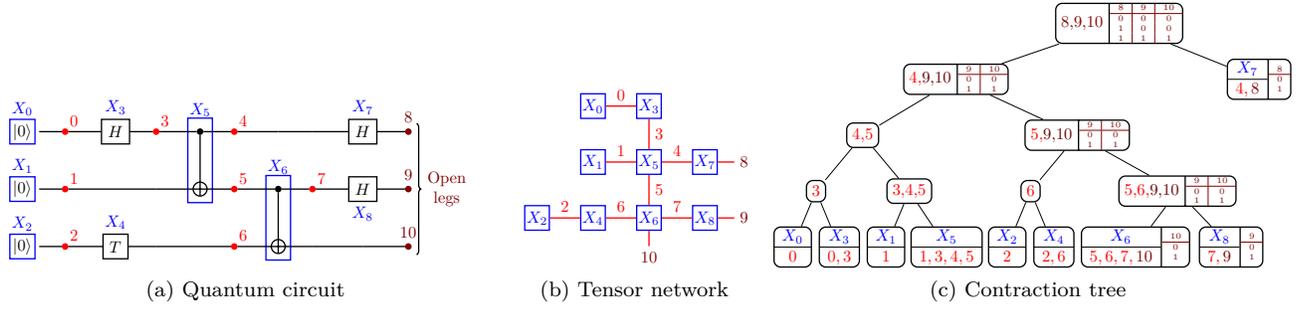

\paragraph{Slicing.} 
In many practical situations, we want to reduce the~memory size used by the~contraction algorithm. 
We can fix some legs in the~tensor network (this is usually also called
\emph{slicing} or \emph{variable projection}~~\cite{Chen:2018, Hyper-optimized:2021, Supremacy:2019}). This way we need to find
the~contraction for all possible values of the~fixed legs (this can be
done in parallel) and then sum up all the results (see Fig.~\ref{fg:slicing}).

Hence in a~tensor network, several legs are fixed to reduce the~maximum
size of intermediate tensors during the~contraction of the~network such
that all intermediate tensors are placed in the~memory of the~device on
which the~contraction is performed.
In the~end, we obtain the~result of the~entire network contraction. This
approach has an~issue: usually the~overall complexity of the~contraction
algorithm increases. However, with a~good optimization, the~loss in the~complexity is not that big.

One particular way to achieve close-to-optimal results is to add to the~list of our local transformation in the~local search method some
additional operations related to slicing. We propose the~following slight modification to 
the~above local search algorithm. We start with the~empty list $S$ of legs used for slicing and every $K$ steps of the~local search method we update $S$ by applying with probability
$1 / 2$ one of the~following two additional steps:
\begin{enumerate}
    \item add to the list $S$ the leg that results in the best memory budget $\mathbf{M}$ reduction;
    \item remove the random leg from $S$.
\end{enumerate}
Let us note that the~objective function during our~local search method requires only a~\emph{local} update on each step, and thus can be implemented very efficiently. However the~same task for these two additional steps usually requires a~\emph{global} update of the~objective function. Nevertheless if $K$ is quite big (e.g., $K=10^{5}$), then we apply these two additional steps not very often,
and the~overall running time of our local search method is almost unchanged.

In the~proposed optimization methods it is very easy to take into account the~implementation details by a~slight modification of the~objective function. We can use this to fine-tune the~contraction tree and slicing obtained by other optimization methods to better fit some particular hardware and software. For example, if there is an~efficiency profile for a~given system, then the~running time $\mathbf{T}$ on this system can be directly estimated, and we can replace $\log_2(\mathbf{C}+\alpha\cdot \mathbf{RW})$ by $\log_2\mathbf{T}$  in the~objective function $f(\mathcal{T})$.

\section{Example of the multi-amplitude algorithm}\label{sc:exapmple}
Let us show the key idea of the above algorithm in a~simple example. Consider a~quantum 
circuit with $3$ qubits (see Fig.~\ref{fg:circuit}(a)) and the~corresponding tensor diagram (see Fig.~\ref{fg:circuit}(b)). In this tensor network diagram $D(X_{0}, \ldots, X_{8})$, for simplicity, we denoted the legs by the~numbers 0--7, and the~open legs by the~numbers 8--10.

If we fix three binary values $s_{1}, s_{2}, s_{3}$ of the output qubits (i.e. we fix the values of the~open legs $8,9,10)$, then we fix the values $T_{0}, \ldots, T_{8}$ of all tensors variables $X_{0}, \ldots, X_{8}$ in the~diagram $D$, and the complex amplitude $\left\langle s_{1} s_{2} s_{3}|C| 0^{n}\right\rangle$ for the bitstring $s_{1} s_{2} s_{3}$ is
equal to the result of the contraction: $\Sigma D(T_{0}, \ldots, T_{8})=\Sigma D(T)$, where $T=(T_0, \ldots, T_8)$.

Now suppose we want to find the complex amplitudes for the following 3-bit strings:
$000,100,111$. For example, to demonstrate our algorithm, we can use the following tree $\mathcal{T}$ given by the contraction expression:
\[{\scriptstyle
\mathcal{T}(X_0, \ldots, X_8) = (((X_0 * X_3) * (X_1 * X_5))*((X_2 * X_4) * (X_6 * X_8))) * X_7
}\]
for the quantum circuit $C$. In order to find our $k=3$ complex amplitudes $\left\langle s_{1} s_{2} s_{3}|C| 0^{n}\right\rangle$
for the bitstrings $s_{1} s_{2} s_{3} \in\{000,100,111\}$ we need to find $\mathcal{T}(T^{1}), \mathcal{T}(T^{2})$, and $\mathcal{T}(T^{3})$,
where each vector of tensors $T^{i}=(T_{0}^{i}, \ldots, T_{8}^{i}), i=1,2,3$, corresponds to our three bit
strings $000,100,111$, respectively. 

In Fig.~\ref{fg:circuit}(c) you can see the~annotated contraction tree
$\mathcal{T}$, where for each internal tree node that corresponds to a~contraction we show the~legs from 0--7 (we sum up over them in this contraction) and the~open legs from 8--10
(the~values of these legs are fixed when we fix the~bitstring $s_1 s_2 s_3$).

For the open legs, we also show their possible values. The number of these values
shows us \emph{how many times} we need to perform the~contraction for this subtree. For
example, for the~subtree $(X_0*X_3)*(X_1*X_5)$ we do not have open legs, hence we need
to calculate it only once when we find $\mathcal{T}(T^1)$, and reuse the result in $\mathcal{T}(T^2)$, and $\mathcal{T}(T^3)$.
At the same time, for the subtree $(X_2*X_4)*(X_6*X_8)$ we have two possible values ($00$
and $11$) for open legs 9 and 10; hence we need to contract this subtree twice. However, if we used $3$ times single-amplitude simulator we would need to contract each subtree
three times.

\section{Details of Multi-tensor algorithm}\label{sc:details}

Here we give a~more detailed description of the multi-tensor algorithm where we want to evaluate a~contraction expression $\cT(X)$ on several tuples of tensors  $T^i=(T_1^i,\dots,T_m^i)$, $i=\overline{1,k}$. 
Let us remind that in this algorithm we have two look-up data structures $\cK_L$ and $\cK_R$ that we call \emph{caches}. Here the keys correspond to the subexpressions of the contraction expression $\cT$, while the values in the cache $\cK_L$ are tensor, and the values in the cache $\cK_R$ are mappings from tuples of integers to tensors. Let us denote by $\cK_R(v;t)$ the value of the mapping $\cK_R(v)$ on the tuple of integers $t$.

If we have a~collection $T=(T^i)_{i=1}^k$ of input arguments, then each variable $X_j$, $j\in[m]$, in the contraction expression $\cT(X_1,\dots,X_m)$ takes finite number of different values. Let $V_j$ be the collection of different tensors that the variable $X_j$ can take, and consider $n_j=|V_j|$, $V=(V_j)_{j=1}^m$. Hence, each tuple $T^i$ is uniquely determined by the tuple of indices $t^i\in\NN^m$ such that $t_j^i\in[n_j]$ and $T^i_j=V_j[t^i_j]$ for all $j\in [m]$. 

The multi-tensor contraction procedure $\mathbf{eval\_all}$ (Algorithm \ref{al:eval-all}) takes the $m$-tuple $V$ of sets of tensors, a~set $t$ of $m$-tuples of indices, and a contraction expression $\cT$. It sorts the set $t$ in the lexicographical order, creates caches $\cK_L$ and $\cK_R$, and sequentially for $i=1,...,m$ calls procedure $\mathbf{eval'}$, which recursively evaluates (reusing temporary results stored in caches) the contraction expression $\cT$ on \emph{one} tuple of tensors defined by $t^i$. The procedure $\mathbf{eval\_all}$ returns the dictionary $R$ where $R(t^i)=\cT(t^i_1,...,t^i_m)$.
In the algorithm it is assumed that the variables of $\cT$ are enumerated in the same order as they occur in  $\cT$.

By $|t|$ we denote the number of elements in the set $t$. If $X_{j_1},...,X_{j_s}$ are the variables of the subexpression $\cT'$ (if $X_{j_i}$ occurs in $\cT'$ earlier than $X_{j_{i'}}$, then $i<i'$), then we will use the following notations:
\begin{equation*}
    t^i_{\cT'}=(t^i_{j_1},...,t^i_{j_s}),\ 
    t_{\cT'}=(t^i_{\cT'})_{i=1}^{|t|},\ 
    V_{\cT'}=(V_{j_1},...,V_{j_s}).
\end{equation*}

\begin{algorithm}[t]
    \tcp{Initialize the caches $\cK_L,\cK_R$ and the~dictionary $R$}
    $\cK_L:=\varnothing$, $\cK_R:=\varnothing$,
    $R:=\varnothing$\;
    Lexicographically sort $t$\;
    \For{$i := 1$ \KwTo $|t|$}{
        $R(t^i):=\mathbf{eval'}(\cT,V,t, i, i+1, \cK_L,\cK_R)$\;
    }
    \Return $R$\;
\caption{Procedure $\mathbf{eval\_all}(\mathcal{T}, V, t)$}
\label{al:eval-all}
\end{algorithm}

\begin{algorithm}[t]
 \lIf{$\cT=X_{j}$}{\Return $V_j[t^i_j]$}
    \eIf{$i=1$ \textbf{or} $t^i_{\cT}\ne t^{i-1}_{\cT}$}{
        Let $\cT=\cT_L*\cT_R$\;
        \If{$\cK_R(\cT)=\mathbf{null}$}{
            \tcp{Call full multi-tensor contraction on the right subtree}
            $\cK_R(\cT):=\mathbf{eval\_all}(\cT_R, V_{\cT_R}, t_{\cT_R})$\;
        }
        \tcp{Recursive call on the left subtree}
        $U_L := \mathbf{eval'}(\cT_L,V,t, i,\max\{i'\,| t^{i'}_{\cT}{=}t^i_{\cT}\}{+}1, \cK_L, \cK_R)$\;
        $U_R := \cK_R(\cT;t^i_{\cT_R})$\;
        \tcp{perform the contraction operation}
        $U:=U_{L} * U_{R}$\;
        \If{$i'\le|t|$ \textbf{and} $t^i_{\cT}=t^{i'}_{\cT}$}{
            \tcp{Store the result in the cache to use it on the next step}
            $\cK_L(\cT):=U$\;
        }
        \lIf{$i'>|t|$}{$\cK_R(\cT)=\mathbf{null}$}
        \Return $U$\;
    }{
        \Return $\mathcal{K}_L(\mathcal{T})$\;
    }
\caption{Procedure$\ \mathbf{eval}'\!(\mathcal{T}, V, t, i, i', \cK_L,\cK_R)$}
\label{al:eval'}
\end{algorithm}

If $\cT=\cT_L*\cT_R$, then $t^i_{\cT}=(t^i_{\cT_L}, {t^i}')$ where ${t^i}'$ is some permutation of $t^i_{\cT_R}$. 
Hence, if $t_{\cT}$ is sorted lexicographically, then $t_{\cT_L}$ is also sorted lexicographically, therefore in each call of $\mathbf{eval'}$ the set $t_{\cT}$ is sorted lexicographically. 
Hence, the result of the evaluation of $\cT$ on the input tuple $t^i_{\cT}$ should be stored in the cache only if $t^{i'}_{\cT}=t^i_{\cT}$ where $i'$ is the value of the parameter~$i$ in the next call of $\mathbf{eval'}$ for the subexpression $\cT$; otherwise $t^j_{\cT}>t^i_{\cT}$ for all $j\ge i'$ and in this case we do not need to store the result in the cache $\cK_L$. 
The~parameter $i'$ for the root expression is equal to $i+1$ since the next call from $\mathbf{eval\_all}$ will be with the parameter $i+1$. 
For the subexpression $\cT_L$ of the expression $\cT=\cT_L*\cT_R$ the parameter $i'$ is the first number after $i$ for which the condition in the second line of Algorithm \ref{al:eval'} is true (i.e. $t^{i'}_{\cT}\ne t^{i'-1}_{\cT}$) or $|t|+1$ if this condition will be always false. Hence we get 
$$i'\!=\min\{i'\mid i'\!>|t|\ \mathrm{or}\ t_{\cT}^{i'}\ne t_{\cT}^i\}=\max\{i'\!\mid t^{i'}_{\cT}=t^i_{\cT}\}+1.$$ 

\paragraph{Complexity estimation.}
For our optimization procedure we need to efficiently estimate the complexity and the memory used in Algorithm \ref{al:eval'}. Suppose we know the sets $V_i$ and the number $k$ of input tuples in $\cT$. Consider a~subexpression $\cT'$ of $\cT$ and its variables $X_{j_1},...,X_{j_s}$. 
This subexpression will be evaluated $|t_{\cT'}|$ times. Since the variable $X_{j_i}$ takes values from $V_{j_i}$, it can take at most $|V_{j_i}|$ different values. Hence the expression $\cT'$ will be evaluated at most $k_{\cT'}=\min\Bigl(\prod_{i=1}^s|V_{j_i}|,k\Bigr)$ times. Therefore, we can estimate the complexity recursively: 
\begin{enumerate}
    \item if $\cT=X_j$, then $k_{\cT}=\min(|V_j|,k)$;
    \item otherwise $\cT=\cT_L*\cT_R$ and 
    $k_{\cT}\le\min(k_{\cT_L}k_{\cT_R}, k)$.
\end{enumerate}
Thus, the values $k_{\cT}$ can be efficiently updated with the complexity $O(1)$ after a~local transformation of the contraction tree.

For the multi-tensor simulation, the contraction cost $\mathbf{C}$ depends not only on the contraction expression $\cT$ but also on the maximal number $k$ of the input tensor tuples. The total contraction cost $\mathbf{C}(\cT, k)$ can be estimated as follows:
\begin{enumerate}
    \item if $\cT=X_j$, then $\mathbf{C}(\cT,k)=0$ (in this case we do not perform the contraction)
    \item otherwise $\cT=\cT_L*\cT_R$, and we have $$\mathbf{C}(\cT,k)=\mathbf{C}(\cT_L,k)+\mathbf{C}(\cT_R,k)+k_{\cT}\mathbf{C}(\cT)$$
    where $\mathcal{C}(\cT)$ is the cost of the contraction in the root of $\cT$ (the contraction of the tensors $\cT_L$ and $\cT_R$), which can be easily calculated from the set of legs of these tensors.
\end{enumerate}
Note that $k_{\cT}$ depends only on the set of variables in the expression $\cT$ and does not depend on the evaluation order of the expression $\cT$. The complexity $\mathbf{C}(\cT_L*\cT_R)$ depends only on the set of output legs of the result of the evaluation of $\cT_L$ and $\cT_R$ and does not depend on the order of the evaluation of $\cT_L$ and $\cT_R$. Hence, when there are some local changes inside the subexpression $\cT$ the complexity gain can be recalculated locally.

When we apply a~local transformation to $\cT=(\cT_1*\cT_2)*\cT_3$ and obtain $\cT'=\cT_1*(\cT_2*\cT_3)$ the subtrees $\cT_1,\cT_2,\cT_3$ remain the same. Taking into account the new subexpression $\cT_2*\cT_3$ and $k_{\cT}=k_{\cT'}$, the difference in the complexity can be calculated as follows:
\begin{multline*}
    \mathbf{C}(\cT,k)-\mathbf{C}(\cT',k)=k_{\cT_1*\cT_2}\mathbf{C}(\cT_1*\cT_2)\\-k_{\cT_2*\cT_3}\mathbf{C}(\cT_2*\cT_3)+k_{\cT}(\mathbf{C}(\cT)-\mathbf{C}(\cT')).
\end{multline*}

\paragraph{Memory estimation.} 
In the previous paragraph, we showed how we can locally recalculate the computational cost of the contraction after a~local transformation in a~contraction tree. However, it is much harder to recalculate the memory size. Below we propose a~fast~algorithm giving us a~heuristic estimate $\mathbf{M}(\cT,k)$ of the memory used in the contraction algorithm $\mathbf{eval\_all}(\cT,V,t)$.

By $m(\cT)$ we denote the size of the contraction result for an~expression $\cT$.
Below we assume that the subexpression $\cT$ is evaluated exactly $k_{\cT}$ times and use the following observations for this case:
\begin{enumerate}
    \item If $k_{\cT}=1$, then $\cT$ is evaluated only once, and hence the cache is not used for storing the subexpression evaluation results.
    \item If it is necessary, then the left and right subexpressions can be swapped without any changes in the complexity.
    \item For each subexpression $\cT=\cT_L*\cT_R$ 
    in the cache $\cK_L$ there is at most one entry, and in the cache $\cK_R$ there are all $k_{\cT_R}$ entries.
    \item The total memory for the temporary results (excluding the caches) is approximately equal to $2\max_{\cT'\subset \cT}m(\cT')$ (in the worst case usually a~large tensor is contracted with a~small tensor, and the result is again a~large tensor of the same shape).
\end{enumerate}
Taking into account the above observations, for an~expression $\cT=\cT_L*\cT_R$ we define:
\begin{align*}
&m_k(\cT)=k_{\cT}m(\cT),\\
&m'(\cT)=\min(m_k(\cT_L)+m(\cT_R), m(\cT_L)+m_k(\cT_R)),\\
&m''(\cT)=\min(m_k(\cT_L), m_k(\cT_R)),\\
&m_{\cK}(\cT)=\begin{cases}
    0 & \mbox{if }k_{\cT}=1, \\
    m'(\cT) & \mbox{if }\min(k_{\cT_L},k_{\cT_R})<k_{\cT}, \\
    m''(\cT) & \mbox{if }k_{\cT_L}=k_{\cT_R}=k_{\cT},
\end{cases}\\
&\mathbf{M}(\cT,k)=\sum_{\cT'\subseteq \cT}m_{\mathcal{K}}(\cT)+2\Bigl(\sum_{\cT'\subseteq \cT}m^p(\cT')\Bigr)^{1/p},
\end{align*}
where $p\ge 1$ is the approximation parameter. Here we approximate the norm $\|\cdot\|_\infty$ by the norm $\|\cdot\|_p$ to make the~target function smooth and locally updatable.

Note that this algorithm does not give the exact memory size. Instead, it calculates only an~approximate value for the optimization procedure. To get the exact memory size one can run Algorithm~\ref{al:eval-all} in an~\emph{emulation mode}, i.e., with a~virtual memory allocator, which only calculates the memory size without any real memory allocation. One can also use dummy tensor contractions where only the tensor shapes are calculated.

\end{document}